\documentclass[pdflatex,sn-mathphys-num]{sn-jnl}

\usepackage{graphicx}%
\usepackage{multirow}%
\usepackage{amsmath,amssymb,amsfonts}%
\usepackage{amsthm}%
\usepackage{mathrsfs}%
\usepackage[title]{appendix}%
\usepackage{xcolor}%
\usepackage{textcomp}%
\usepackage{manyfoot}%
\usepackage{booktabs}%
\usepackage{algorithm}%
\usepackage{algorithmicx}%
\usepackage{algpseudocode}%
\usepackage{listings}%

\usepackage{natbib}

\begin{document}

\title[Article Title]{Learning Deep Phylogeny Through Epistasis-Aware Geometry}


\author[1,4]{\fnm{Jason} \sur{Lamanna}}
\equalcont{These authors contributed equally to this work.}

\author[1,2]{\fnm{Mohammad Erfan} \sur{Mowlaei}}
\equalcont{These authors contributed equally to this work.}


\author[1,2]{\fnm{Xinghua} \sur{Shi}}

\author[1,2,5]{\fnm{Sudhir} \sur{Kumar}}

\author[1,4,5,*]{\fnm{Vincenzo} \sur{Carnevale}}

\affil[1]{\orgdiv{Institute for Genomics and Evolutionary Medicine}, \orgname{Temple University}, \orgaddress{\city{Philadelphia}, \postcode{19122}, \state{PA}, \country{USA}}}

\affil[2]{\orgdiv{Computer and Information Sciences}, \orgname{Temple University}, \orgaddress{\city{Philadelphia}, \postcode{19122}, \state{PA}, \country{USA}}}


\affil[4]{\orgdiv{Institute for Computational Molecular Science}, \orgname{Temple University}, \orgaddress{\city{Philadelphia}, \postcode{19122}, \state{PA}, \country{USA}}}

\affil[5]{\orgdiv{Department of Biology}, \orgname{Temple University}, \orgaddress{\city{Philadelphia}, \postcode{19122}, \state{PA}, \country{USA}}}

\affil[*]{Corresponding author: vincenzo.carnevale@temple.edu}


\abstract{Protein evolution is shaped by epistasis: the effect of a substitution depends on the sequence background in which it occurs. Site-independent substitution models nonetheless support accurate phylogenetic inference, a robustness that plausibly persists only until accumulated substitutions reshape mutational constraints. Potts models describe the epistatic landscapes of protein families but do not reconstruct the histories traversed through them. Here we introduce a generative framework that learns family-specific evolutionary structure directly from sequences evolving on such a landscape. A variational autoencoder is augmented with a covariance-matching term targeting epistatic dependencies and a correlation term tying latent distance to sequence divergence below the point where p-distance saturates. We compare two implementations differing only in latent geometry: EucEvoGen, in Euclidean space, and HyperEvoGen, on a Poincaré ball. Trained on natural protein alignments, both reproduce family-level sequence diversity and, beyond pairwise marginals, recover higher-order statistics more accurately than a Potts-model generator. On sequences evolved along Yule trees under a Potts Hamiltonian, where the true history is known, HyperEvoGen recovers the topology at least as accurately as maximum-likelihood inference, improving on it for four of five tree-distance metrics. Its ancestral reconstructions trail maximum likelihood among closely related sequences but overtake it beyond roughly 0.3 p-distance and retain that advantage as divergence grows. EucEvoGen matches HyperEvoGen generatively but not phylogenetically, showing that generative fidelity alone is insufficient: the global organization of the latent space governs inference at depth. Family-specific representations of epistatically constrained sequence space thus extend phylogenetic inference into the regime where stationary site-independent models lose resolution.}




\maketitle

\section{Introduction}\label{sec1}

Protein families evolve through the accumulation of amino acid substitutions over vast evolutionary timescales. Historical patterns generated by this process provide the foundation for reconstructing evolutionary history, inferring ancestral sequences, estimating divergence times, and understanding the emergence of biological function \cite{nei_molecular_2000}. Since the pioneering work of Dayhoff and the subsequent development of empirical substitution models, molecular evolution has largely been interpreted through stochastic models of amino acid replacement (Dayhoff \cite{dayhoff_model_1978}, JTT \cite{jones_rapid_1992}, WAG \cite{whelan_general_2001}, LG \cite{le_improved_2008}). Their common premise is that evolutionary history can be reconstructed by modeling amino acid substitutions at homologous sites as realizations of a stochastic process whose statistical properties remain sufficiently stable through time. These models have proven remarkably effective across a wide range of phylogenetic applications, often producing accurate evolutionary inferences despite their simplifying assumptions.

Yet decades of theoretical and empirical research have revealed that protein evolution is profoundly influenced by epistasis. \cite{starr_pervasive_2018,shah_contingency_2015} The evolutionary effect of a mutation depends on the sequence background in which it occurs, and substitutions that are favorable in one context may be neutral or deleterious in another. Experimental evolution studies, ancestral protein resurrection experiments, deep mutational scanning, and comparative sequence analyses have collectively demonstrated that contingency, entrenchment, and higher-order interactions are pervasive features of protein evolution.\cite{starr_pervasive_2018,poelwijk_learning_2019}

The coexistence of widespread epistasis and successful phylogenetic inference presents a long-standing paradox. If the effects of mutations depend strongly on sequence background, one might expect stationary substitution models to perform poorly. Yet phylogenetic methods based on such models often recover biologically reasonable evolutionary histories. One possible resolution is that stationary substitution models provide useful approximations over relatively short evolutionary intervals, where the cumulative effects of epistasis remain limited. Indeed, recent studies suggest that phylogenetic inference can remain robust over surprisingly long evolutionary timescales despite substantial violations of model assumptions \cite{magee_robustness_2021,schmelkin_epistasis_2026}.

As proteins diverge over longer evolutionary timescales, however, accumulated substitutions progressively reshape the structural and functional contexts in which future mutations arise. The evolutionary landscape traversed by an evolving sequence can deviate substantially from stationary expectations, as accumulated substitutions alter mutational accessibility, modify selective constraints, and change the evolutionary consequences of subsequent mutations.\cite{di_bari_emergent_2024,shah_contingency_2015} Consequently, the effective evolutionary process itself changes through time. Under these conditions, inference of deep evolution in protein families is likely to be adversely affected by repeated application of the same substitution process over long evolutionary timescales. Evolution becomes increasingly history-dependent, with evolutionary trajectories shaped by continually changing sequence contexts.

A parallel line of research has addressed this challenge by developing statistical models that explicitly incorporate epistatic interactions. Among these, Potts models have emerged as one of the most successful frameworks for describing protein family landscapes. By capturing pairwise dependencies among amino acid positions, Potts models provide powerful descriptions of the constraints governing sequence variation and have yielded important insights into protein structure, fitness landscapes, mutational effects, and evolutionary couplings.\cite{morcos_direct-coupling_2011,figliuzzi_coevolutionary_2016,hopf_mutation_2017} Compared to conventional site-independent substitution models, Potts models provide a substantially richer description of the topography of protein sequence space and explicitly acknowledge that the evolutionary consequences of mutations depend on sequence context.

Yet Potts models address a fundamentally different question from phylogenetic models. Classical substitution models seek to reconstruct evolutionary history, whereas Potts models seek to characterize the likelihood of any given sequence. A substitution model attempts to explain how present-day sequences arose from ancestral states, while a Potts model evaluates the compatibility of sequences with an underlying epistatic landscape. In this sense, substitution models are models of history, whereas Potts models are models of topography. A Potts landscape can identify regions of sequence space favored by evolutionary constraints and estimate the relative probabilities of alternative sequences, but it does not directly reveal how lineages traversed that landscape. Many possible evolutionary routes may connect two extant sequences, and multiple ancestral sequences may be compatible with any two given sequences. Consequently, while Potts models provide a powerful description of where evolution can occur, they do not directly explain how evolution moved through sequence space over time.

This distinction highlights a major conceptual gap in evolutionary modeling. Classical substitution models infer evolutionary history without explicitly representing epistatic landscapes. Potts models represent epistatic landscapes without explicitly reconstructing evolutionary history. Something generally missing between these perspectives is a framework capable of learning evolutionary trajectories through epistatically constrained sequence landscapes. Such a framework would ideally preserve the strengths of both approaches: the ability to represent realistic evolutionary constraints while simultaneously reconstructing the historical paths that connect ancestral and extant sequences.

Here, we introduce a generative framework for learning representations used to infer deep evolutionary trajectories in protein families by harnessing modern representation-learning techniques to capture higher-order dependencies among amino acid positions and learn global relationships among sequences distributed throughout a protein family. Rather than assuming that evolution proceeds through repeated application of a fixed, site-independent substitution process, it learns trajectory structure from sequences evolving across an epistatically constrained landscape.

We examine two complementary implementations of this framework. The first, a modified version of our previously implemented Variational Autoencoder (VAE) \cite{kingma_auto-encoding_2022,mcgee_generative_2021}, represents proteins in a Euclidean latent space, but now uses a loss function augmented with custom terms to induce a relationship between observed sequence divergence and latent space distance. The second, a model we call HyperEvoGen, utilizes a hyperbolic latent space to provide a natural inductive bias for representing hierarchical divergence \cite{nickel_poincare_2017,mathieu_continuous_2019}.

Conventional phylogenetic methods infer trees and ancestral states under a specified substitution process. Our framework shares the objective of reconstructing historical relationships but represents sequence dependencies and evolutionary distances through a learned family-specific latent space. Using the independent site model, the maximum likelihood method identifies ancestral states and trajectories that maximize the probability of the observed sequences under a specified substitution model, while seeking the phylogenetic relationship of the observed sequences \cite{felsenstein_evolutionary_1981,yang_new_1995}. In this process, branch lengths, substitution parameters, and phylogenetic relationships emerge from inferred evolutionary histories linking ancestors to descendants. Our models share this objective but differ in how evolutionary trajectories are represented and inferred. Rather than assuming that all trajectories arise from an independent site model, we learn trajectory structures directly from sequences evolving on epistatically constrained landscapes. In this sense, the phylogenetic tradition extends into regimes where evolutionary dynamics are strongly context dependent and the evolutionary process itself changes as lineages traverse sequence space.

Using evolutionary trajectories generated on Potts-model landscapes, we show that both implementations accurately reproduced family-level sequence diversity and multi-site statistics, but HyperEvoGen produced the strongest phylogenetic and ancestral reconstructions, with results more accurate than state-of-the-art stationary substitution models under conditions of deep epistatic evolution. All outcomes suggest that modeling family-specific evolutionary trajectories through epistatic landscapes offers a powerful extension of classical phylogenetic inference for studying the deep evolutionary history of protein families.

\section{Results}\label{sec2}

Here, we report an integrated evaluation of our evolutionary modeling framework spanning representation learning, phylogenetic inference, and ancestral sequence reconstruction using a curated protein family MSA along with simulated datasets constructed from Potts-model couplings and Yule branching trees. We first obtained an MSA for the Pfam family PF00069, the protein kinase domain \cite{paysan-lafosse_pfam_2025}. We then established the corresponding Potts landscape and trained HyperEvoGen along with our corresponding Euclidean version, which we will now refer to as EucEvoGen. We assessed performance through generative capacity metrics, distance–divergence concordance, phylogenetic tree comparison metrics, and accuracy of ancestral reconstructions. The full experimental pipeline, including preprocessing, simulation, training, and benchmarking, is summarized in the Methods section. To assess whether performance generalizes beyond a single protein family, we repeated training and testing on additional Pfam families: PF00001, PF00005, PF00520, and PF00271, with generative capacity results on these families shown in Supplementary Figures 2-5.

\subsection{EvoGen Architecture}

\begin{figure*}[h!]
\centering
\includegraphics[width=1\textwidth]{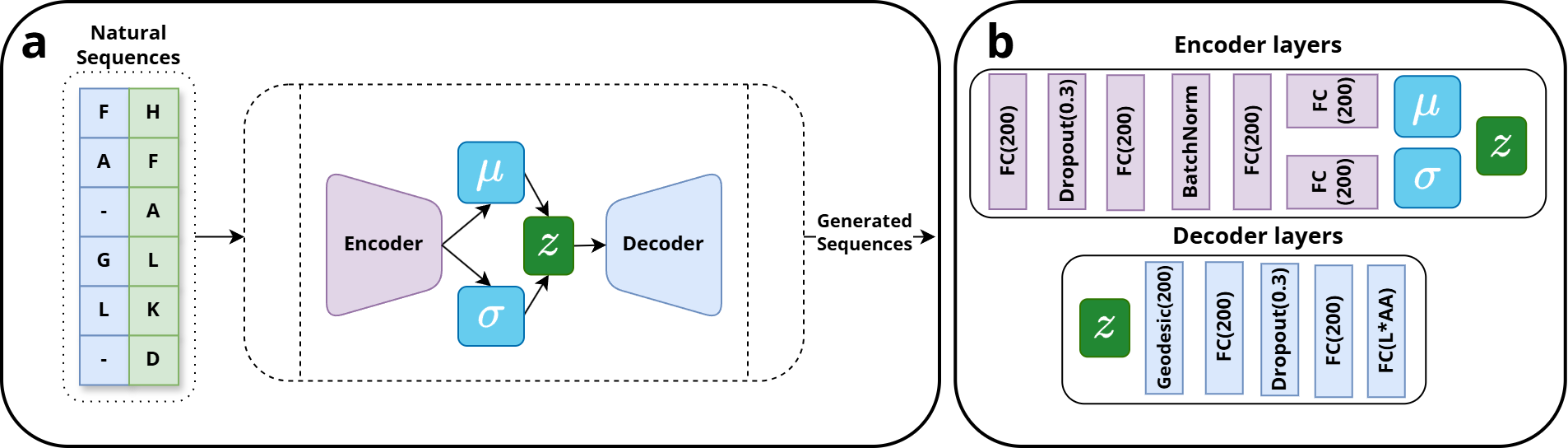} 
\caption{$\vert$ \textbf{Evolutionary modeling framework and HyperEvoGen architecture.} (\textbf{a}) Overall training scheme shared by EucEvoGen and HyperEvoGen. Aligned, one-hot encoded protein sequences are passed through an encoder that parameterizes a latent probability distribution using a mean (\(\mu\)) and variance (\(\sigma\)). A latent vector (\(z\)) is sampled through reparameterization and decoded into a reconstructed protein sequence. The two implementations share this general architecture and training scheme but differ in the geometry of their latent representations and the corresponding operations used for distance calculation, sampling, decoding, and interpolation. (\textbf{b}) Detailed architecture of HyperEvoGen. The encoder contains fully connected layers with 200 hidden units, ELU activations, dropout, and batch normalization, followed by separate linear heads that output \(\mu\) and \(\sigma\). The sampled latent vector is represented in an \(n\)-dimensional Poincaré ball. The decoder passes this representation through fully connected layers and a geometry-aware geodesic layer before producing an \(L \times AA\)-dimensional vector of amino acid logits, where \(L\) is the sequence length and \(AA\) is the number of possible amino acid states. In contrast to HyperEvoGen, EucEvoGen performs the corresponding latent-space and decoding operations using Euclidean geometry and therefore does not require the hyperbolic geodesic layer.
}
\label{fig:Arch}
\end{figure*}

Our evolutionary modeling framework is implemented using two variational autoencoder (VAE) architectures that differ primarily in their latent geometry: EucEvoGen uses an \(n\)-dimensional Euclidean latent space, whereas HyperEvoGen uses an \(n\)-dimensional Poincaré ball \cite{kingma_auto-encoding_2022,nickel_poincare_2017,mathieu_continuous_2019}. Both implementations consist of an encoder, a continuous latent representation, and a decoder (Fig.~\ref{fig:Arch}). The encoder maps one-hot encoded protein sequences to the parameters of a latent probability distribution, from which latent vectors are obtained through reparameterized sampling. In EucEvoGen, sampling and distance calculations are performed using Euclidean operations. In HyperEvoGen, the corresponding operations are defined according to the geometry of the Poincaré ball. The decoder maps sampled latent vectors back to logits over the possible amino acid states at each sequence position. The HyperEvoGen decoder includes a geodesic layer \cite{mathieu_continuous_2019} that operates on the Poincaré latent representation according to the manifold geometry before the output amino acid logits are calculated. EucEvoGen instead decodes directly from its Euclidean latent representation using standard fully connected operations. The encoder and decoder are constructed from fully connected layers with ELU activations \cite{clevert_fast_2016} and use dropout \cite{srivastava_dropout_2014} and batch normalization \cite{ioffe_batch_2015} to promote generalization and training stability. Spectral normalization \cite{miyato_spectral_2018} is applied to the fully connected layers throughout the architecture.

The models were implemented in Python using PyTorch \cite{paszke_pytorch_2019}, with Pythae \cite{chadebec_pythae_2022} serving as the primary library for the VAE architecture and latent-distribution operations. Unless otherwise specified, the analyses reported here used a five-dimensional latent space, a learning rate of \(2\times10^{-5}\), and 3,500 training epochs. A batch size of 400 was used for models trained on natural protein sequences, whereas a batch size of 1,000 was used for models trained on synthetic sequences generated from the evolutionary simulations. The covariance-loss weight was scaled according to the number of sequence elements in each training batch and was set to \(\lambda_{\mathrm{cov}}=B\times L\), where \(B\) is the batch size and \(L\) is the aligned sequence length. For HyperEvoGen, the Poincaré ball curvature was fixed at \(c=1.0\). Its latent prior was parameterized as a Riemannian normal distribution \cite{mathieu_continuous_2019}, and its approximate posterior was represented by the wrapped normal distribution \cite{nagano_wrapped_2019} implemented in Pythae. The shared architecture and training objectives encourage both models to recover family-specific sequence constraints, while their alternative latent geometries define different distance functions and interpolation paths for phylogenetic inference and ancestral sequence reconstruction.

\begin{figure*}[h]
\centering
\includegraphics[width=0.9\textwidth]{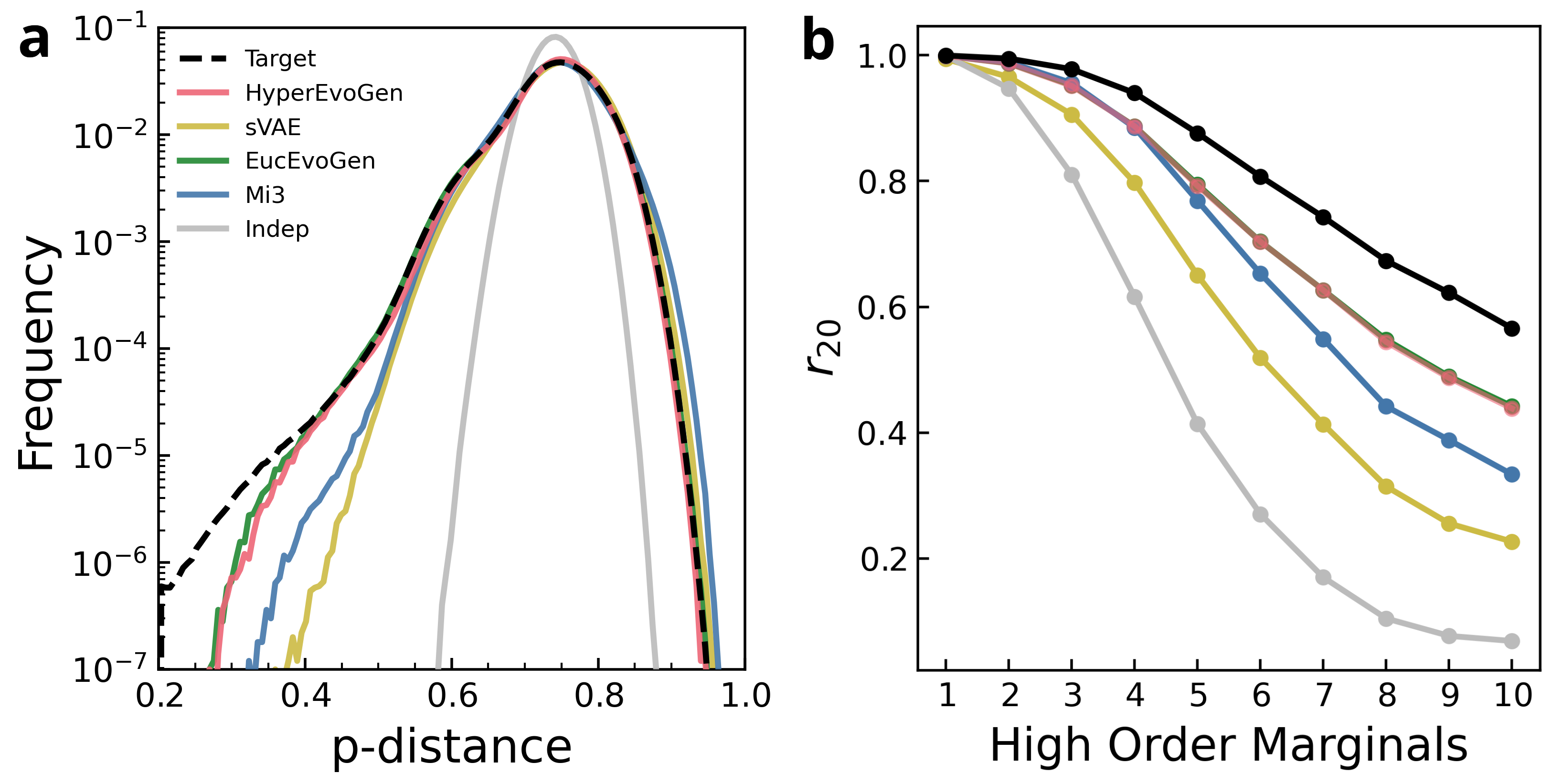} 
\caption{$\vert$ \textbf{HyperEvoGen and EucEvoGen show improved generative recovery of sequence diversity and higher-order statistics.} Here, we evaluate the generative capacity of HyperEvoGen and EucEvoGen against Mi3, a Potts-model-based sequence generator \cite{haldane_mi3-gpu_2021}, and sVAE, an earlier iteration of this work \cite{mcgee_generative_2021}. (\textbf{a}) P-distance frequency distributions are shown with a logarithmically scaled y-axis to highlight differences in the distribution tails. Although all models reproduce the most frequently observed p-distances, HyperEvoGen and EucEvoGen more accurately recover the infrequent distances found at the tails, demonstrating comparable performance across the two latent geometries. (\textbf{b}) The $r_{20}$ score \cite{mcgee_generative_2021} measures the Pearson correlation between the frequencies of the 20 most common non-contiguous amino acid "words" of orders 1--10 in generated and target sequences. Supplementary Figure 8 provides a more detailed view of the frequency correlations used to calculate $r_{20}$. HyperEvoGen and EucEvoGen achieve similar performance and outperform the other models, particularly for higher-order marginals, indicating that their shared training objectives recover complex sequence statistics across both latent space geometries. Because Mi3 is explicitly designed to reproduce observed single-site frequencies and pairwise covariances, it retains a slight advantage for the order-1 and order-2 marginals.}
\label{fig:GenCap}
\end{figure*}

\subsection{Generative Capacity}
We evaluated the framework's generative capacity, the models' ability to generate new sequences from the distributions learned during training \cite{mcgee_generative_2021,sgarbossa_generative_2023}, by comparing generated samples with a target MSA held out from the reference dataset. We used two complementary diagnostics: the distribution of pairwise p-distances, which quantifies overall sequence diversity, and higher-order marginal statistics calculated from noncontiguous amino acid  “words”, which assess the recovery of increasingly complex multi-site sequence patterns. Together, these metrics evaluate whether the models reproduce both the diversity and statistical dependencies present in the target protein family. 10,000 sequences generated by EucEvoGen and HyperEvoGen are compared to those from a target MSA of 10,000 natural PF00069 sequences and reproduced complex multi-site statistics, with closely matched performance across the two latent geometries. These results indicate that the shared training objectives enable both models to learn family-specific patterns of sequence variation. (Fig.~\ref{fig:GenCap}).

\subsection{Phylogenetic Inference}
We simulated large-scale evolutionary datasets using a Metropolis–Hastings framework \cite{metropolis_equation_1953,hastings_monte_1970} in which proposed mutations were accepted or rejected according to changes in Hamiltonian energy derived from Potts-model couplings \cite{de_la_paz_epistatic_2020,bisardi_modeling_2022,di_bari_emergent_2024}. Evolution began from a single ancestral sequence and proceeded along branching structures generated under a Yule process \cite{yule_iimathematical_1997}, modeling lineage splitting as a continuous-time stochastic process. From a given Yule tree, four major clades are selected, defined by the first three root splits. Within each clade, pairs of tips were sampled across divergence levels measured by p-distance, spanning the full range from closely related to highly diverged sequences. Training sets for the model were created by withholding these evaluation sequences from the full dataset, ensuring that HyperEvoGen and EucEvoGen were both tested on held-out tip sequences.

\begin{figure*}[ht]
\centering
\includegraphics[width=1\textwidth]{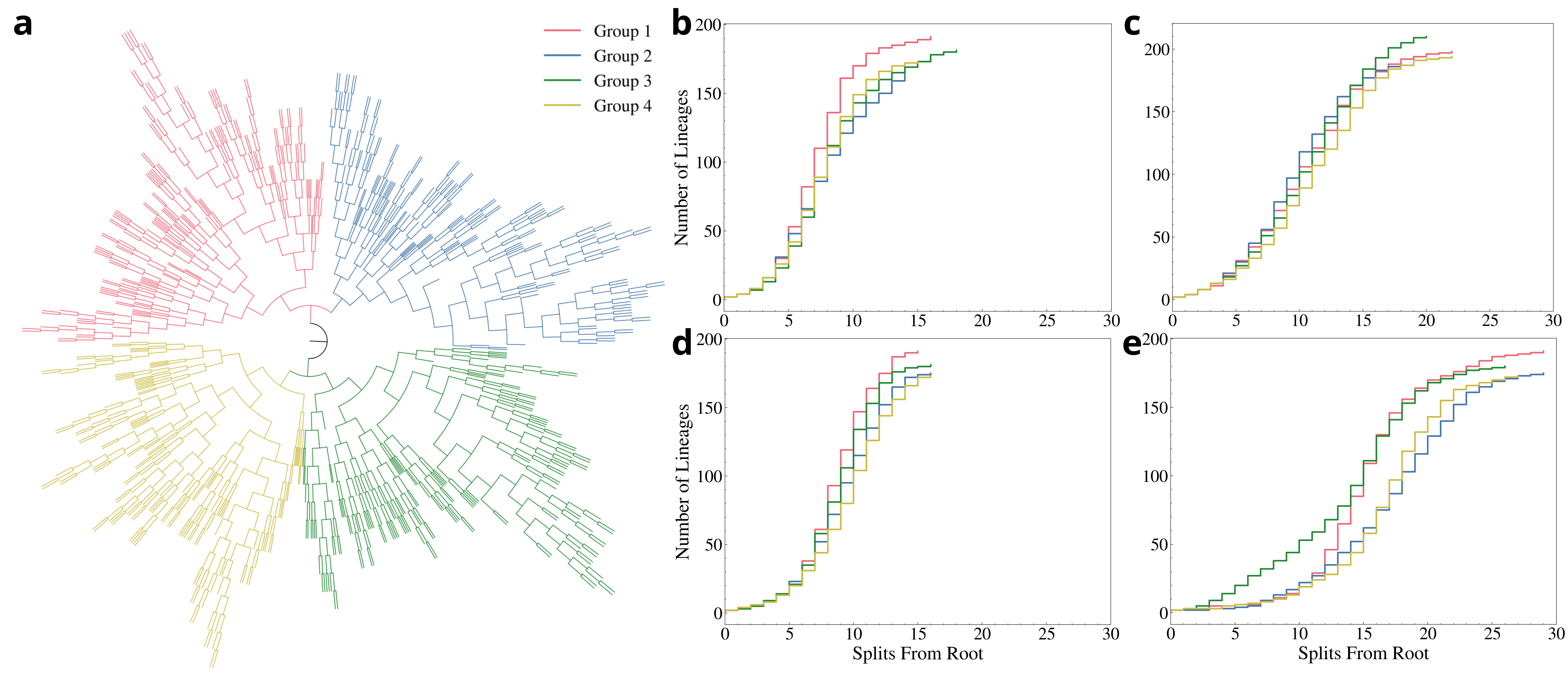} 
\caption{$\vert$ \textbf{Lineage-through-time analysis compares diversification patterns among ground-truth, EucEvoGen, HyperEvoGen, and IQ-TREE phylogenies.} (\textbf{a}) Ground-truth phylogeny used for benchmarking, shown as a radial tree with the four major clades highlighted by color (Groups 1--4). (\textbf{b}) Lineage-through-time (LTT) curves calculated from the shown ground-truth tree. Curves show the number of extant lineages as a function of the number of splits from the root for each clade. (\textbf{c}) Corresponding LTT curves for the EucEvoGen-inferred tree. (\textbf{d}) LTT curves for the HyperEvoGen-inferred tree. (\textbf{e}) LTT curves for an IQ-TREE reconstruction generated from the same tip sequences. Both EucEvoGen and HyperEvoGen show closer agreement with the ground-truth lineage-through-depth profiles in this analysis, although direct topology metrics reveal substantially poorer overall reconstruction by EucEvoGen.}
\label{fig:PhyloInf}
\end{figure*}

For phylogenetic reconstruction, all tip sequences were embedded in the respective latent space of each model, and pairwise distances were calculated to generate Neighbor-Joining trees \cite{saitou_neighbor-joining_1987}. The distances from EucEvoGen are measured simply with Euclidean distance, however, in the case of HyperEvoGen, these distances correspond to geodesics on the n-dimensional Poincaré ball. The generated trees were compared against trees built from conventional p-distances and maximum-likelihood inference using IQ-TREE \cite{wong_iq-tree_2025,kalyaanamoorthy_modelfinder_2017}. Trees reconstructed from HyperEvoGen use a custom hybrid distance metric, explained in detail later in the Methods section, which combines normalized p-distances with scaled latent space geodesics and better preserve diversification rates than those made with IQ-TREE or embedding distance on their own. This hybrid distance method improves HyperEvoGen but does not comparably improve EucEvoGen as shown in Supplementary Figure 9, indicating that the gain cannot be explained by the inclusion of the p-distance mixture alone, but instead relies more on the inductive bias introduced by HyperEvoGen’s hyperbolic representation. Lineage-through-time (LTT) curves \cite{nee_tempo_1992} derived from hybrid-distance trees closely matched those from the ground-truth Yule simulations, showing minimal distortion (Fig. ~\ref{fig:PhyloInf}).

To more comprehensively quantify phylogenetic reconstruction accuracy, we compared each inferred tree with the ground-truth topology using several complementary tree-distance metrics. Normalized Robinson--Foulds distance (nRF) \cite{robinson_comparison_1981} measures disagreement in the set of internal bipartitions between two trees, providing a direct assessment of exact split recovery. Because nRF treats each split as either matching or non-matching, we additionally evaluated quartet distance \cite{estabrook_comparison_1985}, which measures disagreement among the topological relationships of four-taxon subsets and provides a more local assessment of tree structure. Matching split distance (MSD) \cite{bogdanowicz_matching_2012} further relaxes the requirement for exact split identity by optimally matching similar bipartitions between trees, while clustering information distance (CID) \cite{smith_information_2020} quantifies differences between splits according to their shared information content. Finally, topological path difference \cite{steel_distributions_1993} measures discrepancies in the number of branches separating pairs of taxa across the two trees. For each of these metrics, smaller values indicate greater similarity to the ground-truth phylogeny. The results for all of these measures can be found in \ref{tab:tree}

\begin{table*}[b]
\centering
\begin{minipage}{\textwidth}
\centering

\begin{tabular*}{\textwidth}{@{\extracolsep{\fill}}c|ccccc}
\hline
\textbf{Method} & \textbf{nRF} & \textbf{Quartet} & \textbf{MSD} & \textbf{CID} & \textbf{Path} \\
\hline
HyperEvoGen & \textbf{0.5416} & 0.3108 & \textbf{0.2967} & \textbf{0.4096} & \textbf{3016.37} \\
EucEvoGen & 0.7989 & 0.4927 & 0.4558 & 0.6017 & 5445.89 \\
IQ-TREE & 0.5501 & \textbf{0.3040} & 0.3295 & 0.4445 & 4231.34 \\
\hline
\end{tabular*}

\caption{$\vert$ \textbf{HyperEvoGen most accurately recovers the ground-truth phylogenetic topology across complementary tree-distance metrics.} Inferred phylogenies were compared with the ground-truth tree using normalized Robinson--Foulds distance (nRF), quartet distance, matching split distance (MSD), clustering information distance (CID), and topological path difference. HyperEvoGen achieves the lowest error across all metrics except quartet distance, outperforming EucEvoGen and IQ-TREE on the remaining metrics. Its quartet distance is slightly higher than that of IQ-TREE. Values are equally weighted means across five tip sets sampled from the same simulated tree. Lower values indicate better agreement with the ground truth.}
\label{tab:tree} 
\end{minipage}
\end{table*}

\subsubsection{Hierarchical Topology Recovery}

\begin{figure*}[h!]
\centering
\includegraphics[width=1\textwidth]{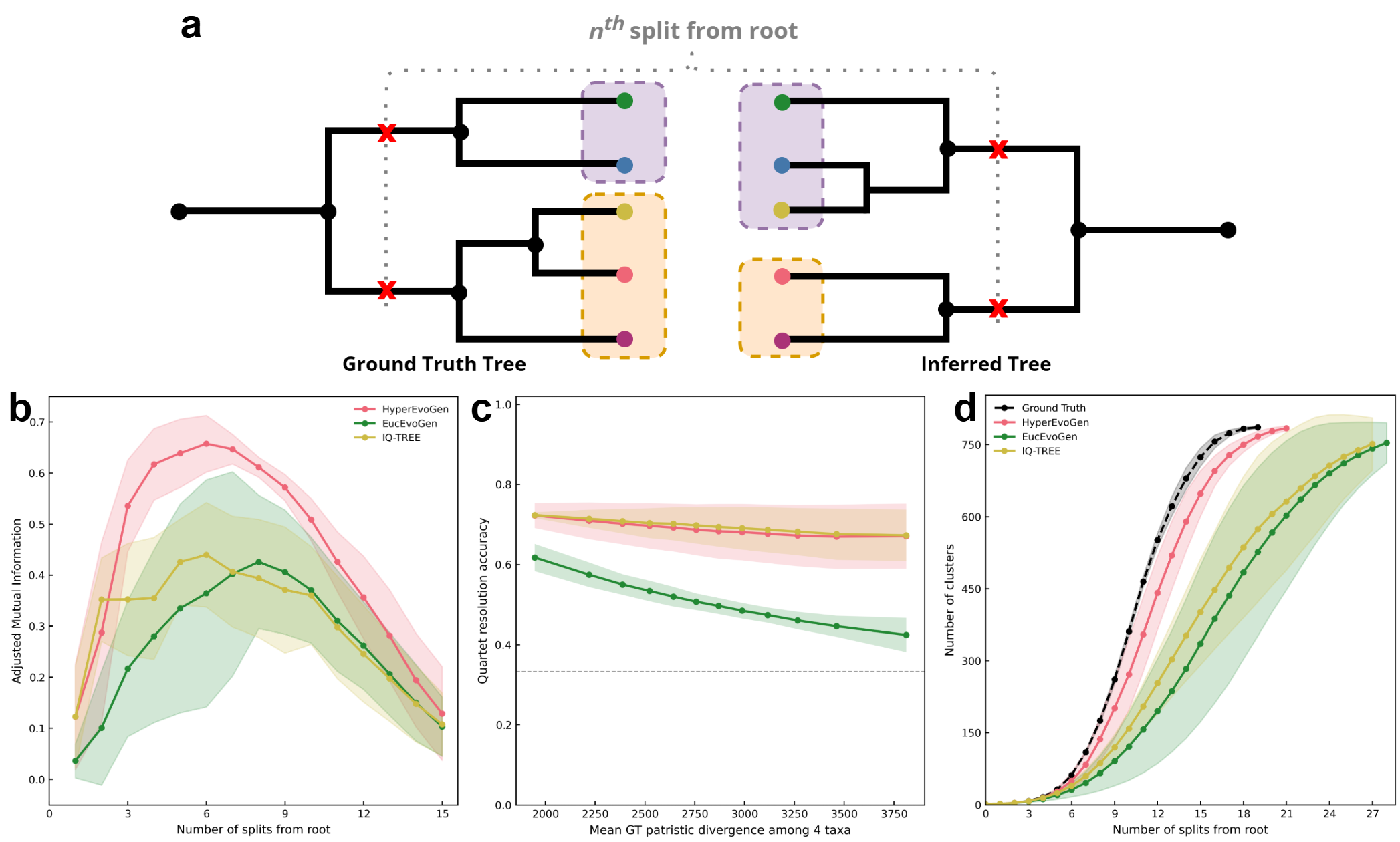} 
\caption{$\vert$ \textbf{Multiscale comparisons reveal scale-dependent recovery of ground-truth phylogenetic structure.} (\textbf{a}) Comparison scheme. Ground-truth and inferred trees are partitioned at the same topological depth, measured as the number of splits from the root, and the resulting tip assignments are compared. (\textbf{b}) Adjusted mutual information (AMI) between partitions as a function of split depth. HyperEvoGen shows the highest mean agreement across most evaluated depths. (\textbf{c}) Quartet resolution accuracy versus mean ground-truth patristic divergence among the four taxa. HyperEvoGen and IQ-TREE show similar mean accuracy across the evaluated range, whereas EucEvoGen exhibits lower accuracy and a stronger decline with divergence. The dashed line indicates chance accuracy ($1/3$). (\textbf{d}) Number of clusters at each split depth. HyperEvoGen most closely follows the ground-truth profile, whereas EucEvoGen and IQ-TREE require greater depths to resolve comparable numbers of clusters. In (\textbf{b–d}), lines show equally weighted means across five tip sets sampled from the same simulated tree; shaded regions indicate $\pm$1 sample standard deviation across datasets.}
\label{fig:AMI}
\end{figure*}

To determine whether these differences in global tree accuracy were concentrated at particular hierarchical scales, we next evaluated topology agreement as a function of split depth using adjusted mutual information (AMI) \cite{vinh_information_2010}. While the global tree-distance metrics are effective in summarizing the disagreement across the entire topology, AMI-by-depth measures how well inferred trees preserve the organization of taxa into clades at progressively finer levels of the hierarchy.(Fig. ~\ref{fig:AMI}). Shallow depths quantify recovery of broad, deep clade structure, whereas larger depths quantify agreement among progressively finer subclades. The theoretical upper bound is \(\mathrm{AMI}=1\) at all evaluated depths.

We used AMI-by-depth as a follow up topology comparison because global split-based metrics can obscure scale-dependent agreement. Robinson--Foulds distances penalize discordant bipartitions individually and can report large distances even when major clade organization is preserved \cite{smith_information_2020}. In contrast, AMI-by-depth evaluates whether tips remain grouped into similar clade partitions at each hierarchical level, allowing disagreement in broad tree structure to be distinguished from disagreement in fine-scale local topology. 

At very deep split depths, the ground-truth partition may approach a singleton partition in which each tip forms its own cluster. In this regime, AMI becomes strict, with high AMI requiring the inferred tree to isolate the same individual tips at the same depth, whereas remaining multi-tip clusters produce little agreement beyond chance.

\subsection{Ancestral Sequence Reconstruction}

After training, HyperEvoGen and EucEvoGen were used to reconstruct ancestral sequences for the sampled tip sets through a shared latent-space reconstruction procedure. The two implementations differ in the geometry of their latent representations and, consequently, in the definitions of distance and interpolation used during reconstruction. For HyperEvoGen, tip sequences were embedded by the encoder into an \(n\)-dimensional Poincaré ball, and pairwise distances were calculated as hyperbolic geodesic distances. For EucEvoGen, the sequences were embedded into an \(n\)-dimensional Euclidean latent space, and pairwise distances were calculated using the standard Euclidean metric. For each model, the resulting pairwise distance matrix was supplied to the Neighbor-Joining algorithm \cite{saitou_neighbor-joining_1987} to construct a phylogenetic tree, hereafter referred to as the model's embedding tree. In testing, we discovered that mixing HyperEvoGen derived distances with p-distances according to a sigmoid function improved its performance. 

Ancestral sequences were inferred by traversing each embedding tree in post-order. At each internal node, the latent representations of its two child nodes were obtained and the path connecting them was calculated according to the geometry of the corresponding model. HyperEvoGen uses the geodesic connecting the child embeddings within the Poincaré ball, whereas EucEvoGen uses the straight-line path connecting the child embeddings in Euclidean space. One hundred evenly spaced points were sampled along the respective path, and the midpoint embedding was passed to the model's decoder to generate the amino acid sequence assigned to the ancestral node. The reconstructed state at each internal node was then used in subsequent reconstruction steps toward the root, allowing ancestral sequences to be inferred recursively throughout the tree.

Neither implementation uses branch-length information to weight the placement of ancestral states, nor does either assume an explicit amino acid substitution model during latent-space reconstruction. Instead, ancestral states are positioned between their descendant states according to the distance and interpolation operations defined by each model's latent geometry. This shared procedure provides a generative alternative to conventional maximum-likelihood ASR: HyperEvoGen performs interpolation along hyperbolic geodesics, whereas EucEvoGen performs the equivalent interpolation along Euclidean line segments \cite{liberles_ancestral_2007,sennett_extant_2024,horta_ancestral_2022,de_leonardis_reconstruction_2025}.

To benchmark ASR accuracy, ancestral sequences reconstructed by HyperEvoGen and EucEvoGen were compared with maximum-likelihood reconstructions produced by IQ-TREE \cite{wong_iq-tree_2025} using the same sampled tip sets. The best-fitting amino acid substitution model was selected automatically using ModelFinder \cite{kalyaanamoorthy_modelfinder_2017}, and phylogenetic inference was performed under the selected model. This provided a conventional substitution-model baseline in which alignment sites are treated as conditionally independent given the tree and model parameters. Ancestral states were reconstructed using IQ-TREE's \texttt{-asr} option under the inferred tree and selected substitution model \cite{yang_new_1995}. The resulting ancestral sequences were used for comparison with the latent-space reconstructions. IQ-TREE phylogenies were midpoint rooted \cite{farris_estimating_1972} for root-dependent evaluations while for HyperEvoGen and EucEvoGen, the root was retained at the final joining node returned by the distance-based tree-construction procedure. This algorithmic root was used to orient the inferred hierarchy and does not constitute a separate estimate of ancestral root placement. 

\begin{figure*}[h!]
\centering
\includegraphics[width=1\textwidth]{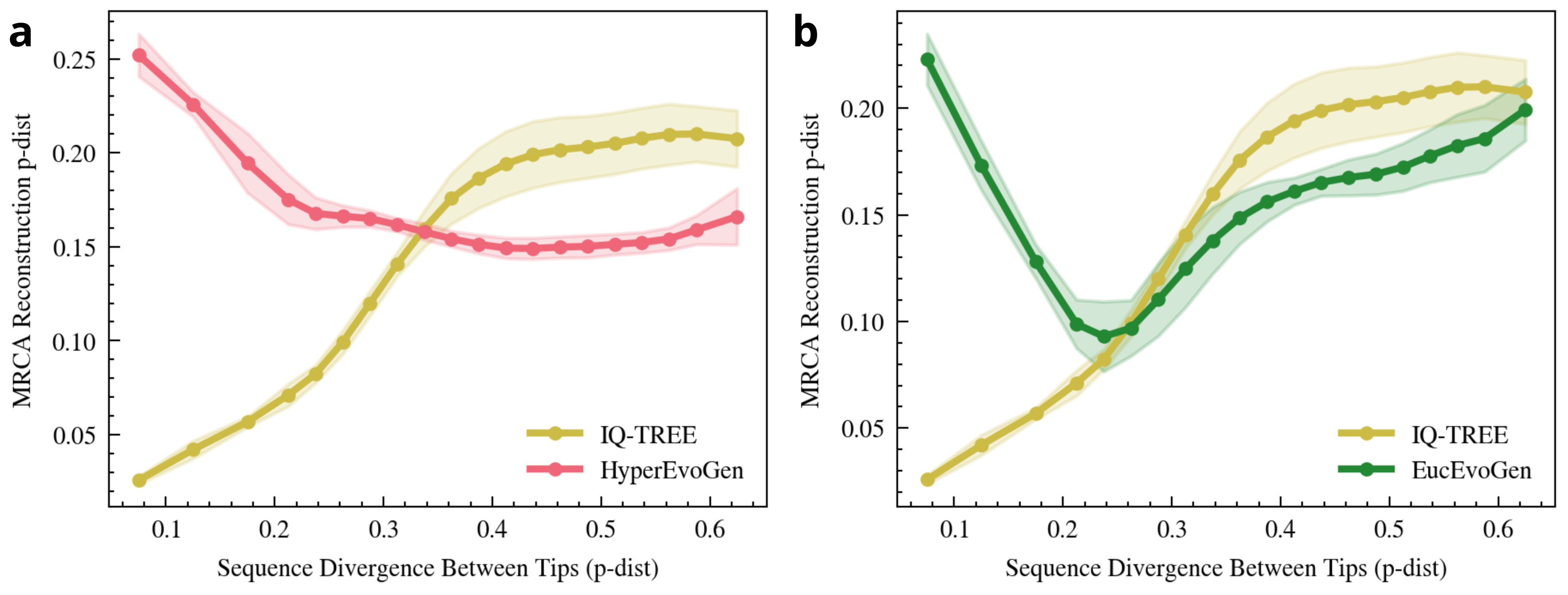} 
\caption{$\vert$ \textbf{MRCA reconstruction accuracy as a function of tip--tip sequence divergence.} We evaluated ancestral sequence reconstruction at the most recent common ancestor (MRCA) of pairs of tips grouped by their observed tip--tip sequence divergence (p-distance). Reconstruction error was measured as the p-distance between the inferred and ground-truth MRCA sequences, with lower values indicating greater accuracy. Lines show equally weighted means of the dataset-specific bin means across five tip sets sampled from the same simulated tree. Shaded regions indicate ±1 sample standard deviation across these five means. (\textbf{a}) Mean MRCA reconstruction error for IQ-TREE and HyperEvoGen. IQ-TREE performs better at low sequence divergence, whereas HyperEvoGen becomes more accurate at moderate divergence and maintains the advantage as divergence increases. (\textbf{b}) Mean MRCA reconstruction error for IQ-TREE and EucEvoGen. EucEvoGen also achieved lower mean error across much of the moderate-to-high divergence range, but its error increased with divergence and approached the IQ-TREE curve in the highest-divergence bins.}
\label{fig:MRCA}
\end{figure*}

Reconstruction error at the most recent common ancestor (MRCA) was evaluated as a function of observed tip--tip sequence divergence using both the p-distance between the reconstructed and ground-truth MRCA sequences and the difference in their Potts Hamiltonian energies. HyperEvoGen exhibited a characteristic crossover in reconstruction performance: IQ-TREE was more accurate at very low tip divergence, whereas HyperEvoGen overtook IQ-TREE at approximately \(0.3\) p-distance and maintained lower reconstruction error as divergence increased (Fig.~\ref{fig:MRCA}a). EucEvoGen also performed less accurately than IQ-TREE at low divergence and became comparatively more accurate across part of the moderate-divergence range, but its reconstruction error subsequently increased at greater divergence and approached that of IQ-TREE (Fig.~\ref{fig:MRCA}b). These results indicate that the shared latent-space reconstruction framework can become advantageous as closely related sitewise signals weaken, while HyperEvoGen's geodesic interpolation provides the more consistent improvement across moderate-to-deep evolutionary divergence.

\section{Discussion}

This work introduces a generative framework for learning evolutionary trajectories through epistatically constrained protein sequence landscapes. Classical substitution models provide representations of evolutionary histories but generally simplify or omit dependencies among sites, whereas Potts models describe the topography of epistatic sequence landscapes without directly reconstructing the historical paths taken through them. Our framework seeks to bridge these perspectives by learning a continuous representation that captures family-specific sequence constraints while organizing proteins according to evolutionary divergence. Using Potts-based evolution along Yule trees with known histories, we evaluated two implementations of this framework, EucEvoGen and HyperEvoGen, across sequence generation, evolutionary distance estimation, phylogenetic reconstruction, and ancestral sequence reconstruction. The results demonstrate that the shared generative objectives enable both models to recover important statistical properties of the underlying protein family, while the choice of latent geometry substantially affects how effectively this information is translated into deep evolutionary inference.

Both EucEvoGen and HyperEvoGen reproduced the overall sequence diversity and multi-site statistics of the target protein family. Their comparable performance in the generative-capacity benchmarks indicates that realistic sequence generation does not depend exclusively on hyperbolic geometry. Instead, this arises from the shared architecture and training objectives, with both models using a covariance-based loss term to encourage recovery of dependencies among amino acid positions and a divergence-based loss to relate local sequence divergence to latent distance. The recovery of higher-order marginal statistics is particularly notable because the covariance term directly constrains only pairwise relationships. Thus, higher-order agreement appears to emerge from the learned sequence distribution rather than being explicitly specified by the covariance objective. These statistical results should not be interpreted as proof that every generated sequence is functional or that the models recover all forms of biological epistasis. Rather, they show that the models reproduce family-specific patterns of coordinated sequence variation beyond independent single-site frequencies.

The comparison between EucEvoGen and HyperEvoGen clarifies the role of latent geometry within this framework. Both implementations use the same general trajectory-learning and epistasis-aware objectives, but they parameterize evolutionary relationships differently. EucEvoGen represents proteins in a Euclidean latent space, whereas HyperEvoGen uses an \(n\)-dimensional Poincaré ball in which available volume grows exponentially with distance from the origin. This property provides a natural inductive bias for representing branching and hierarchical relationships \cite{nickel_poincare_2017,mathieu_continuous_2019,matsumoto_novel_2021,macaulay_fidelity_2023,jiang_learning_2022}. The ability of EucEvoGen to match HyperEvoGen in generative-capacity testing demonstrates that negative curvature is not required to learn the statistical distribution of a protein family. Its less consistent performance in phylogenetic and deep ancestral reconstruction, however, suggests that learning realistic sequence statistics and organizing those statistics into an evolutionarily informative global representation are distinct challenges.

HyperEvoGen more consistently preserved evolutionary relationships across large divergence ranges, including regions in which direct sequence dissimilarity became increasingly compressed by repeated substitutions. Its latent distances supported stronger recovery of deep phylogenetic structure and more accurate ancestral reconstruction than the Euclidean implementation. This result is consistent with the expectation that hyperbolic space can represent tree-like expansion with less distortion than an unconstrained Euclidean space \cite{sarkar_low_2011,sala_representation_2018}. However, the comparison does not establish that Euclidean representations are intrinsically incapable of encoding deep evolutionary histories. EucEvoGen recovered realistic sequence distributions and showed comparable MRCA reconstruction accuracy relative to IQ-TREE across portions of the moderate-to-deep divergence range. The results instead indicate that, under the implementations evaluated here, hyperbolic geometry provides a more effective inductive bias for converting the shared trajectory-learning objectives into a globally organized evolutionary representation.

The ancestral reconstruction results further illustrate how the relative strengths of these approaches change with evolutionary depth. Conventional maximum-likelihood ancestral sequence reconstruction estimates ancestral states under a specified substitution process, commonly treating sites as conditionally independent given the tree and model \cite{liberles_ancestral_2007,sennett_extant_2024,thornton_resurrecting_2004,horta_ancestral_2022,ferreiro_trends_2026,ferreiro_substitution_2024}. At shallow divergence, where observed substitutions retain strong local signal and repeated changes remain limited, IQ-TREE generally reconstructed ancestral sequences more accurately than either latent-space model. However, as divergence increased, its reconstruction error rose, whereas HyperEvoGen became more accurate and maintained lower mean reconstruction error across more increasingly diverged sequences.

Conceptually, these findings are consistent with the history-dependent nature of protein evolution. The consequences of a mutation depend on the sequence background in which it occurs, and accumulated substitutions can alter the accessibility and effect of subsequent changes \cite{domingo_causes_2019,di_bari_emergent_2024,nishikawa_epistasis_2021,figliuzzi_coevolutionary_2016,hopf_mutation_2017,muniz-trejo_epistatic_2025}. Stationary independent-site models can remain effective over shallow and even moderately long evolutionary intervals, as reflected by the strong performance of IQ-TREE at low divergence. Their approximations may become less informative as repeated substitutions accumulate and lineages traverse changing sequence contexts. By learning from sequences evolved under an epistatic Potts landscape, the present framework represents coordinated sequence variation and uses that information when estimating relationships among proteins. The advantage of HyperEvoGen within high divergence ranges supports the central hypothesis that trajectory-aware, family-specific representations may be especially useful for studying deep evolutionary relationships.

Our approach to ancestral reconstruction should nevertheless be interpreted as an estimator of evolutionary trajectories rather than a direct recovery of the precise historical path followed by a lineage. HyperEvoGen places candidate ancestors along geodesics between descendant embeddings and decodes selected latent positions into amino acid sequences. EucEvoGen performs an analogous operation in Euclidean space. Because the decoders are trained on the family-level sequence distribution, these interpolations are intended to remain within regions supported by the learned representation rather than independently selecting the most probable state at each site. However, neither latent interpolation nor statistical agreement with a target alignment guarantees that a reconstructed sequence is structurally stable, functional, or historically accessible. Experimental resurrection \cite{thornton_resurrecting_2004}, structural characterization, and comparison with alternative ancestral posterior distributions will be required to establish the biological realism of the inferred trajectories.

This study remains intentionally grounded in a controlled benchmark. Most quantitative evaluations used sequences evolved under a Potts model inferred from the PF00069 protein kinase family and propagated along Yule trees, providing known topologies, branch lengths, substitution histories, and ancestral sequences. This design isolates the effects of epistatic sequence constraints and enables direct evaluation of evolutionary inference. Potts models are approximations of protein-family landscapes and primarily encode pairwise statistical dependencies. Natural protein evolution additionally involves processes not represented here, including insertions and deletions, recombination, domain rearrangement, lineage-specific rate variation, shifting environmental pressures, and changes in function. Performance on these simulations therefore demonstrates that trajectory-learning models can exploit epistatic structure under known conditions, but it does not establish equivalent accuracy across all natural protein families or evolutionary regimes.

Several components of the current framework also remain heuristic. Latent dimensionality, hyperbolic curvature, the relative weights of the reconstruction, covariance, and divergence objectives, and the parameters used to combine observed and embedding-based distances were selected empirically. These settings may not transfer optimally among protein families or divergence ranges. The contrast between EucEvoGen and HyperEvoGen makes this limitation particularly important: poor performance of one Euclidean configuration should not be interpreted as a definitive comparison between all Euclidean and hyperbolic representations. Systematic optimization of model capacity, geometry, loss weighting, and distance calibration will be necessary to determine which advantages arise from curvature itself and which can be recovered through alternative objectives or mappings in Euclidean space.

Several extensions present themselves from the current limitations. Learned distances could be incorporated into likelihood or Bayesian phylogenetic frameworks as family-specific priors, branch-length models, or proposal distributions rather than being used only to construct distance-based trees. Combining the framework with protein language models \cite{rives_biological_2021,lupo_protein_2022} or structure-aware networks may improve its ability to represent higher-order constraints that are not captured by pairwise Potts statistics. Extending the evolutionary simulations and model architecture to accommodate indels, recombination, changing selective environments, and multidomain proteins would provide more realistic tests of generalization. Finally, experimental resurrection of ancestors on which EucEvoGen, HyperEvoGen, and conventional methods disagree would offer the most direct test of whether trajectory-based reconstruction yields proteins with more plausible stability, activity, and specificity.

In summary, this study presents a framework that connects models of epistatic sequence-space topography with the reconstruction of evolutionary history. EucEvoGen and HyperEvoGen demonstrate that biologically informed generative objectives can recover sequence diversity and complex multi-site statistics across different latent geometries. Their differing performance on phylogenetic and ancestral reconstruction further shows that generative fidelity alone is not sufficient for deep evolutionary inference: the organization of the latent representation also matters. In the present implementation, hyperbolic geometry provides the more effective representation of deep, branching evolutionary relationships, while the Euclidean model demonstrates that several benefits of the framework arise from the shared training objectives rather than curvature alone. Together, these findings support family-specific trajectory learning as a promising extension of classical phylogenetic inference and establish a foundation for evolutionary models that integrate epistatic constraints, learned distance structure, and generative ancestral reconstruction.

\section{Methods}\label{sec3}

\subsection{Preprocessing}

A multiple sequence alignment (MSA) of the Pfam protein kinase family PF00069 \cite{paysan-lafosse_pfam_2025} comprising over 830,000 sequences was used as the dataset for the results shown thus far. The alignment was generated using a profile hidden Markov model (HMM) \cite{eddy_accelerated_2011} and spans 3,131 positions, with approximately 60\% gap characters overall. Alignment columns which are 80\% gap characters or more were removed to eliminate poorly aligned positions, reducing the alignment length to 167 positions. Sequences sharing more than 80\% amino acid identity were also iteratively removed to reduce redundancy, yielding a non-redundant alignment of ~143,000 sequences. The curated alignment was then divided into training, validation, and test subsets, with 100,294 sequences in the training set, 14,328 in the validation set, and the test set reduced to a total of 10,000 sequences for metric computation time. These processed training alignments were used as input for Potts model coupling inference (using the Mi3-GPU software \cite{haldane_mi3-gpu_2021}) and for training the HyperEvoGen and EucEvoGen models.

\subsection{Potts Model Inference}

Potts model couplings were inferred from the preprocessed MSA using the Mi3-GPU algorithm \cite{haldane_mi3-gpu_2021}. The Potts model parameters are comprised of a local field term $h_i(a)$ for each position $i$ and amino acid $a$, and a coupling matrix $J_{ij}(a, b)$ for each pair of positions $i$ and $j$ \cite{levy_potts_2017}. The sequence alphabet was restricted to the 20 standard amino acids plus a gap character, and coupling matrices were symmetrized such that $J_{ij}(a, b) = J_{ji}(b, a)$. The Hamiltonian (statistical energy) of a sequence $s$ was defined as

\begin{equation}
H(s) = \sum_i h_i(s_i) + \sum_{i<j} J_{ij}(s_i, s_j).
\label{eq:hamiltonian}
\end{equation}

These energy values were used to simulate sequence evolution via a Metropolis–Hastings Monte Carlo procedure, enforcing the learned coevolutionary constraints during mutation and selection.

\begin{figure*}[b]
\centering
\includegraphics[width=1.0\textwidth]{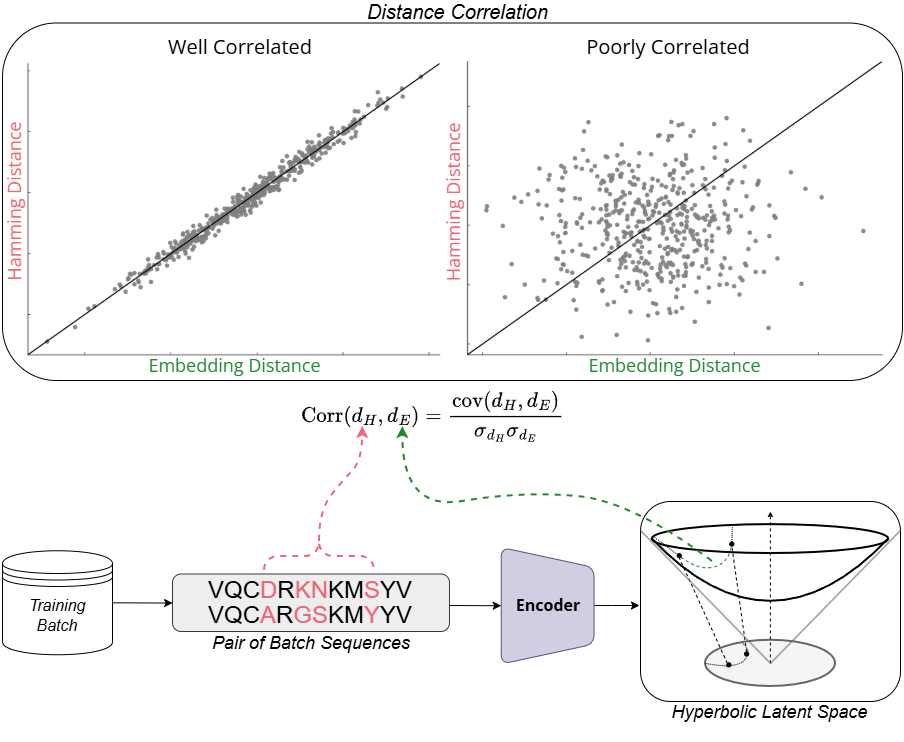} 
\caption{$\vert$ \textbf{Pairwise sequence and embedding distances are aligned through a correlation-based loss term.} Here we illustrate the distance-correlation term in the compound loss. For each training mini-batch, pairs of sequences above the designated cutoff are taken and their Hamming distances in sequence space ($d_H$) are compared with the corresponding geodesic distances between their embeddings ($d_E$) in the hyperbolic latent space produced by the encoder. The top panels show examples of possible outcomes: a “Well Correlated” case and a “Poorly Correlated” case. The distance-correlation loss explicitly encourages the former behavior by maximizing the Pearson correlation over all pairwise distances in the batch. The loss term is computed by taking sequences from a training batch and calculating pairwise Hamming and embedding distance matrices. The resulting correlation between these matrices is then used as a regularizing loss term to encourage the latent geometry to preserve evolutionary divergence.}
\label{fig:Corr}
\end{figure*}

\subsection{Compound Loss Function}

Model training uses a compound loss function that combines the standard VAE objective with two additional biologically motivated terms. The VAE’s base objective is the negative evidence lower bound (ELBO), consisting of a sequence reconstruction log-likelihood and a Kullback–Leibler divergence term that regularizes the latent distribution \cite{kingma_auto-encoding_2022}. These additional terms are a covariance matching loss and a divergence correlation loss. 

The covariance-based term is included to capture epistatic dependencies between residue positions \cite{levy_potts_2017,hopf_mutation_2017,de_juan_emerging_2013,domingo_causes_2019,figliuzzi_coevolutionary_2016}. For each training minibatch, the covariance matrix of amino acid frequencies across alignment positions is computed for both the real sequences in the batch and the VAE’s reconstructed sequences; the model then minimizes the mean squared error between these two covariance matrices. This covariance matching loss penalizes the model if it fails to reproduce the pairwise co-occurrence patterns present in real protein sequences, thereby biasing the VAE to allocate latent capacity to epistatic interactions.

\begin{equation}
\mathcal{L}_{\mathrm{cov}}
=
\frac{1}{L^{2}}
\sum_{i=1}^{L}
\sum_{j=1}^{L}
\left(
\mathrm{Cov}_{\mathrm{gen}}[i,j]
-
\mathrm{Cov}_{\mathrm{nat}}[i,j]
\right)^{2}.
\label{eq:cov_loss}
\end{equation}
The second additional term is a divergence correlation loss that aligns distances in the latent space with evolutionary distances observed in sequence space within a low-to-moderate divergence regime where $p$-distance remains informative. To avoid training on pairs where $p$-distance becomes unreliable due to saturation at higher divergence, we apply a $p$-distance cutoff and compute the correlation only over pairs below this threshold. For each batch, pairwise $p$-distances (the fraction of mismatched amino acids between two sequences) \cite{nei_molecular_2000} are calculated between input sequences and compared to pairwise distances within their corresponding latent embeddings. The loss is then defined as one minus the Pearson correlation between the two distance matrices, penalizing the VAE when latent distances do not correlate with sequence divergences below the threshold.

\begin{equation}
\mathcal{L}_{\mathrm{corr}}
=
1 - \mathrm{Corr}(d_H, d_E)
=
1 - \frac{\mathrm{Cov}(d_H, d_E)}{\sigma_{d_H}\,\sigma_{d_E}}.
\label{eq:corr_loss}
\end{equation}

This term drives the encoder to place similar sequences close together in space and more divergent sequences farther apart, up to the point where raw sequence distances become saturated (Fig. ~\ref{fig:Corr}). The total training objective is the sum of all these components: the VAE reconstruction–KL loss (negative ELBO), the covariance loss, and the correlation loss. By optimizing this combined loss, HyperEvoGen and EucEvoGen learn latent representations that preserve evolutionary relationships and sequence realism alike.

To quantify the contribution of each biologically motivated component, we performed an ablation analysis in which the models were retrained with individual terms removed from the compound objective (i.e., excluding the covariance matching term or the divergence-correlation term), isolating how each loss shapes sequence realism, epistatic signal recovery, and the alignment between latent geometry and evolutionary divergence. The models were tested with a p-distance cutoff of 0.3 and without, results for which are shown in Supplementary Figure 7.

\subsection{Evolutionary Distances from Hyperbolic Embeddings}
All distances in the HyperEvoGen latent space are computed as geodesic distances under the hyperbolic metric of the Poincaré ball. Each latent embedding $z$ is constrained to lie within the unit ball (reflecting the negative curvature model), and the distance between two latent points is given by the geodesic length in this space rather than a Euclidean distance \cite{nickel_poincare_2017,mathieu_continuous_2019}.

\begin{equation}
d(z,y)
=
\frac{1}{\sqrt{c}}\,
\cosh^{-1}\!\left(
1
+
2c\,\frac{\lVert z-y\rVert^{2}}
{\left(1-c\lVert z\rVert^{2}\right)\left(1-c\lVert y\rVert^{2}\right)}
\right).
\label{eq:poincare_distance}
\end{equation}

This hyperbolic distance $d_H$ grows rapidly as points approach the boundary of the ball, which enables the model to represent large evolutionary divergences without compression. In practice, the analytic Poincaré distance formula is used to calculate pairwise distances between latent vectors, ensuring the computed distances reflect the underlying geometry. These geodesic distances are utilized during training (for example, in the correlation loss to link latent distances with $p$-distances) and in downstream analyses. For phylogenetic inference, the learned pairwise distances between sequence embeddings are treated as an evolutionary distance matrix. Applying Neighbor-Joining to the latent distance matrix yields a tree that can be directly compared to traditional trees built from sequence-based distances \cite{saitou_neighbor-joining_1987}. This approach to distance computation in the latent space is critical for capturing hierarchical relationships and ensures that the model’s continuous embeddings serve as a biologically meaningful metric for evolutionary divergence.

\subsection{Hybrid distances for phylogenetic reconstruction}

\begin{figure*}[t]
\centering
\includegraphics[width=1.0\textwidth]{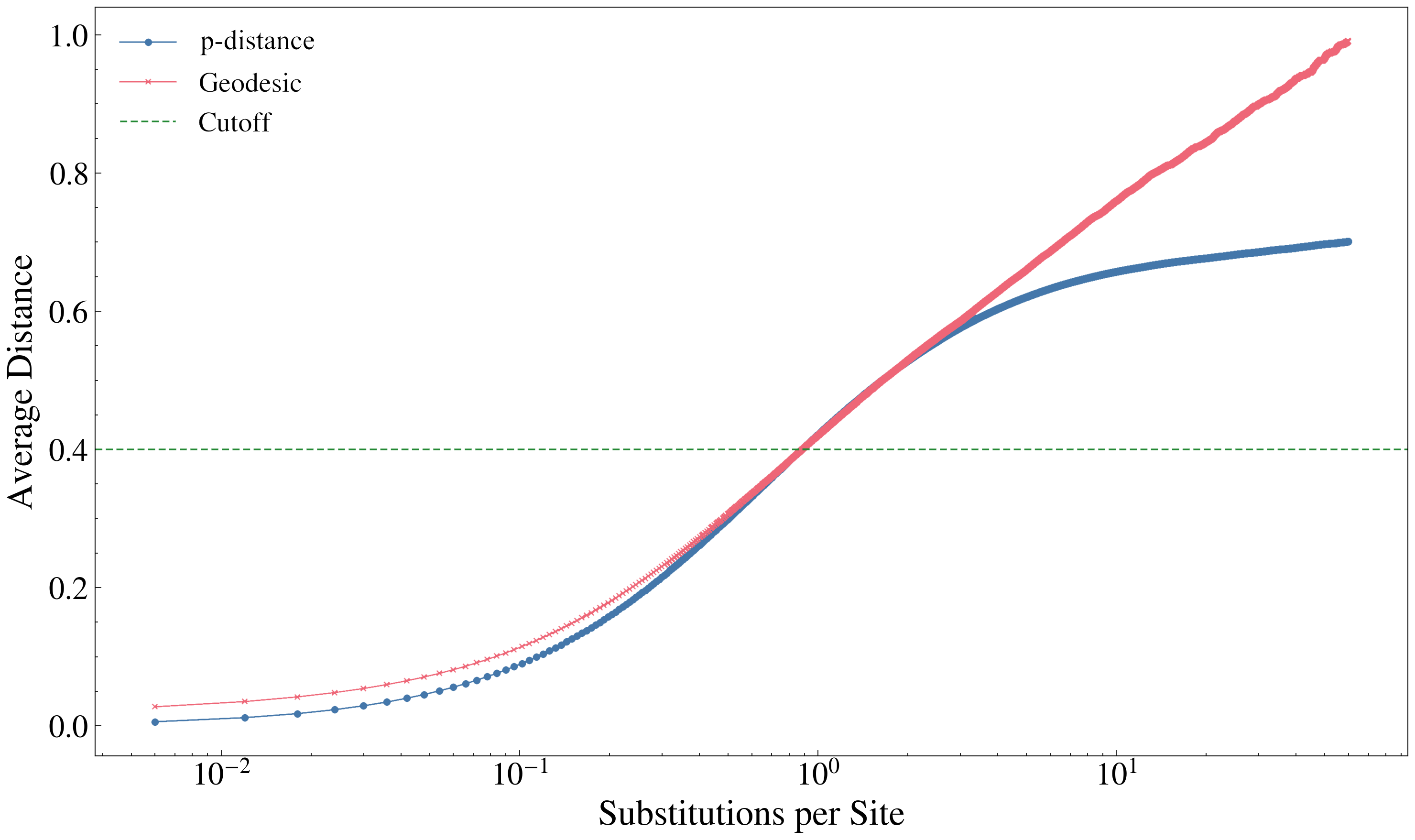} 
\caption{$\vert$ \textbf{Geodesic distance preserves signal at high divergence.} Average pairwise distance is plotted as a function of substitutions per site for both p-distance (blue) and HyperEvoGen’s latent-space geodesic distance (red). The x-axis is logarithmically scaled, such that equal horizontal intervals represent multiplicative rather than additive increases in evolutionary divergence. This expands the low-divergence region while accommodating the broad range of high-divergence values. At low divergence, the two measures follow similar trends, indicating that local neighborhoods in sequence space are well approximated by p-distance. As divergence increases, the geodesic distance continues to grow while p-distance begins to saturate. The horizontal dashed line marks the p-distance cutoff, used later for our hybrid distance method, chosen because it lies near the transition between the two regimes.}
\label{fig:Dist}
\end{figure*}

To combine observed sequence differences at low divergence with learned latent distances at greater divergence, we constructed a hybrid pairwise distance matrix. For sequences \(i\) and \(j\), the observed p-distance was calculated as \(p_{ij}=H_{ij}/L\), where \(H_{ij}\) is the number of mismatched alignment positions and \(L\) is the aligned sequence length. The corresponding embedding distance, \(e_{ij}\), was calculated using the hyperbolic geodesic metric for HyperEvoGen or the Euclidean metric for EucEvoGen.

Because embedding distances and p-distances have different scales, embedding distances were rescaled using sequence pairs within a window around the transition point \(b\). Specifically, we defined

$$
\mathcal{W}=\{(i,j):i<j,\;b-\delta\leq p_{ij}\leq b+\delta\},
\qquad
\bar{e}_{\mathcal{W}}=
\frac{1}{|\mathcal{W}|}\sum_{(i,j)\in\mathcal{W}}e_{ij},
$$

and calculated

$$
\widetilde{e}_{ij}=\frac{b}{\bar{e}_{\mathcal{W}}}e_{ij}.
$$

This normalization sets the mean rescaled embedding distance within the window to \(b\), providing a common local scale for combining the two distance measures. The scaling factor was calculated separately for each input distance matrix.

The contribution of the embedding distance was determined by a sigmoid function of observed p-distance:

$$
w_{ij}=\frac{1}{1+\exp[-a(p_{ij}-b)]}.
$$

The hybrid distance was then defined as

$$
d^{\mathrm{hybrid}}_{ij}
=(1-w_{ij})p_{ij}+w_{ij}\widetilde{e}_{ij}.
$$

We used \(a=10\), \(b=0.4\), and \(\delta=0.2\), giving a normalization window of \(0.2\leq p_{ij}\leq0.6\). Embedding distances were rescaled so that their mean within this window equaled \(0.4\). The parameter \(a\) controls the steepness of the sigmoid transition, while \(b\) specifies the p-distance at which the observed and rescaled embedding distances receive equal weight. Consequently, p-distance contributes more strongly below \(0.4\), whereas embedding distance contributes more strongly above \(0.4\), with a smooth transition between the two. The normalization window determines the scaling factor and does not impose a hard boundary on the weighting function.

The resulting hybrid distance matrix was supplied to Neighbor-Joining for phylogenetic reconstruction. This construction is intended to retain the information provided by observed sequence differences among closely related sequences while incorporating learned distance structure where p-distance becomes less informative. The scaling aligns the two components locally; it does not calibrate hybrid distances in substitutions per site. Additional information about the hybrid-distance analysis is provided in Supplementary Figure 10.

\subsection{Adjusted Mutual Information}
To compare inferred phylogenies to the ground-truth tree across hierarchical scales, we use adjusted mutual information (AMI) \cite{vinh_information_2010} computed on depth-resolved clade partitions. At each split depth \(d\), defined as the number of topological splits from the root, each rooted tree is converted into a partition of shared tips. Each tip is assigned to the clade reached after \(d\) splits from the root and if a terminal tip is reached before depth \(d\), that tip is retained as its own terminal cluster. This procedure produces a ground-truth partition \(U^{(d)}\) and an inferred-tree partition \(V^{(d)}\), both defined over the same set of \(N\) tips.

Given two partitions \(U\) and \(V\), let \(n_{ij}\) denote the number of tips shared between cluster \(i\) in \(U\) and cluster \(j\) in \(V\). Let \(a_i = \sum_j n_{ij}\) and \(b_j = \sum_i n_{ij}\) denote the corresponding marginal cluster sizes. The mutual information between the two partitions is

\begin{equation}
I(U,V)
=
\sum_i \sum_j
\frac{n_{ij}}{N}
\log
\left(
\frac{N n_{ij}}{a_i b_j}
\right),
\end{equation}

where terms with \(n_{ij}=0\) are omitted. Mutual information measures how much knowing a tip's cluster assignment in one tree reduces uncertainty about its cluster assignment in the other tree. However, raw mutual information is biased by the number and size of clusters, and can increase as partitions become more highly resolved even in the absence of meaningful topological agreement.

To correct for this effect, we used adjusted mutual information. The entropy of partition \(U\) is defined as:

\begin{equation}
H(U)
=
-\sum_i
\frac{a_i}{N}
\log
\left(
\frac{a_i}{N}
\right),
\end{equation}

with \(H(V)\) defined analogously. AMI subtracts the expected mutual information between random partitions with the same cluster-size structure:

\begin{equation}
\mathrm{AMI}(U,V)
=
\frac{
I(U,V) - \mathbb{E}[I(U,V)]
}{
\frac{1}{2}\left[H(U)+H(V)\right] - \mathbb{E}[I(U,V)]
}.
\end{equation}

Here, \(\mathbb{E}[I(U,V)]\) is the expected mutual information under a random partitioning model that preserves the marginal cluster sizes. Under this normalization, \(\mathrm{AMI}=1\) indicates identical partitions, whereas \(\mathrm{AMI}\approx 0\) indicates agreement no greater than expected by chance. AMI can also compare partitions with different numbers of clusters, which is useful because two trees cut at the same split depth may produce different numbers of clades.

\subsection{Evolutionary Simulations}

Phylogenetic trees were simulated under a Yule branching process \cite{yule_iimathematical_1997}, each containing 100,000 extant sequences (tips). Two sets of trees were generated corresponding to different total divergence depths, with root-to-tip distances of approximately 3 substitutions per site. Individual branch lengths were drawn from an exponential distribution and converted into a discrete number of mutation steps, allowing the continuous-time Yule process to be used with a discrete, step-based simulation framework.

Sequence evolution along these model trees was performed using a Metropolis–Hastings Monte Carlo algorithm \cite{metropolis_equation_1953,hastings_monte_1970} under the Potts-model Hamiltonian energy function. Starting from a single root sequence, point mutations were proposed along each branch and accepted or rejected according to the Metropolis–Hastings acceptance criterion based on the change in Potts energy, incorporating coevolutionary constraints into the evolutionary trajectory. This procedure yields a full set of evolved protein sequences at the tree’s tips, after which any identical tip sequences were identified and removed to eliminate redundancy.

These simulated trees are partitioned into four major clades based on the first three splits at the root. Within each clade, all pairwise p-distances are computed and binned from the smallest observed divergence to the largest. Tip pairs were then randomly sampled from each divergence bin. To evaluate sensitivity to tip sampling, we repeated the divergence-stratified sampling procedure five times on the same simulated tree containing 100,000 tips, producing five evaluation datasets with different sampled tip sets. These datasets share an underlying evolutionary history and do not represent independent evolutionary simulations. For the five-dataset AMI, MRCA, tree topology metrics, and multiscale analyses, statistics were calculated separately for each dataset before aggregation. Binned curves show the arithmetic mean of the five dataset-specific bin means and shaded regions indicate the mean plus or minus one sample standard deviation across the five dataset-specific values. For multiscale analyses, pooled ground-truth divergence values were used where applicable to define shared bin boundaries; performance statistics were subsequently calculated within each dataset and averaged with equal weight. For AMI, values were calculated separately at each split depth and then averaged across datasets. 


All data used in this study are publicly available. Multiple sequence alignments for Pfam families (including PF00069) were obtained directly from InterPro \cite{blum_interpro_2025}, which hosts Pfam family annotations and alignments. Source data and processed datasets generated during this study are provided with this paper and/or its Supplementary Information.

\section*{Code availability}
The source code of HyperEvoGen will be publicly available on GitHub \url{https://github.com/jclamanna/HyperEvoGen}) upon paper acceptance.


\bibliography{Prelim,Prelim_additions}%

@article{de_juan_emerging_2013,
  title     = {Emerging methods in protein co-evolution},
  volume    = {14},
  copyright = {2013 Springer Nature Limited},
  issn      = {1471-0064},
  url       = {https://www.nature.com/articles/nrg3414},
  doi       = {10.1038/nrg3414},
  language  = {en},
  number    = {4},
  journal   = {Nature Reviews Genetics},
  publisher = {Nature Publishing Group},
  author    = {{de Juan}, David and Pazos, Florencio and Valencia, Alfonso},
  month     = apr,
  year      = {2013},
  pages     = {249--261}
}

@article{de_leonardis_reconstruction_2025,
  title    = {Reconstruction of {Ancestral} {Protein} {Sequences} {Using} {Autoregressive} {Generative} {Models}},
  volume   = {42},
  issn     = {1537-1719},
  url      = {https://doi.org/10.1093/molbev/msaf070},
  doi      = {10.1093/molbev/msaf070},
  number   = {4},
  journal  = {Molecular Biology and Evolution},
  author   = {De Leonardis, Matteo and Pagnani, Andrea and Barrat-Charlaix, Pierre},
  month    = apr,
  year     = {2025},
  pages    = {msaf070}
}

@article{di_bari_emergent_2024,
  title     = {Emergent time scales of epistasis in protein evolution},
  volume    = {121},
  url       = {https://www.pnas.org/doi/10.1073/pnas.2406807121},
  doi       = {10.1073/pnas.2406807121},
  number    = {40},
  journal   = {Proceedings of the National Academy of Sciences},
  publisher = {Proceedings of the National Academy of Sciences},
  author    = {Di Bari, Leonardo and Bisardi, Matteo and Cotogno, Sabrina and Weigt, Martin and Zamponi, Francesco},
  month     = oct,
  year      = {2024},
  pages     = {e2406807121}
}

@article{domingo_causes_2019,
  title    = {The {Causes} and {Consequences} of {Genetic} {Interactions} ({Epistasis})},
  volume   = {20},
  issn     = {1545-293X},
  doi      = {10.1146/annurev-genom-083118-014857},
  language = {eng},
  journal  = {Annual Review of Genomics and Human Genetics},
  author   = {Domingo, Júlia and Baeza-Centurion, Pablo and Lehner, Ben},
  month    = aug,
  year     = {2019},
  pages    = {433--460}
}

@article{ferreiro_substitution_2024,
  title    = {Substitution {Models} of {Protein} {Evolution} with {Selection} on {Enzymatic} {Activity}},
  volume   = {41},
  issn     = {0737-4038},
  url      = {https://pmc.ncbi.nlm.nih.gov/articles/PMC10873502/},
  doi      = {10.1093/molbev/msae026},
  number   = {2},
  journal  = {Molecular Biology and Evolution},
  author   = {Ferreiro, David and Khalil, Ruqaiya and Sousa, Sergio F and Arenas, Miguel},
  month    = feb,
  year     = {2024},
  pages    = {msae026}
}

@article{ferreiro_trends_2026,
  title    = {Trends in substitution models of protein evolution for phylogenetic inference},
  volume   = {214},
  issn     = {1055-7903},
  url      = {https://www.sciencedirect.com/science/article/pii/S1055790325001903},
  doi      = {10.1016/j.ympev.2025.108473},
  journal  = {Molecular Phylogenetics and Evolution},
  author   = {Ferreiro, David and Pazos, Elena and Arenas, Miguel},
  month    = jan,
  year     = {2026},
  pages    = {108473}
}

@article{figliuzzi_coevolutionary_2016,
  title    = {Coevolutionary {Landscape} {Inference} and the {Context}-{Dependence} of {Mutations} in {Beta}-{Lactamase} {TEM}-1},
  volume   = {33},
  issn     = {0737-4038},
  url      = {https://www.ncbi.nlm.nih.gov/pmc/articles/PMC4693977/},
  doi      = {10.1093/molbev/msv211},
  number   = {1},
  journal  = {Molecular Biology and Evolution},
  author   = {Figliuzzi, Matteo and Jacquier, Hervé and Schug, Alexander and Tenaillon, Oliver and Weigt, Martin},
  month    = jan,
  year     = {2016},
  pages    = {268--280}
}

@article{haldane_mi3-gpu_2021,
  title      = {Mi3-{GPU}: {MCMC}-based inverse {Ising} inference on {GPUs} for protein covariation analysis},
  volume     = {260},
  issn       = {0010-4655},
  shorttitle = {Mi3-{GPU}},
  url        = {https://www.sciencedirect.com/science/article/pii/S0010465520301193},
  doi        = {10.1016/j.cpc.2020.107312},
  journal    = {Computer Physics Communications},
  author     = {Haldane, Allan and Levy, Ronald M.},
  month      = mar,
  year       = {2021},
  pages      = {107312}
}

@article{hopf_mutation_2017,
  title    = {Mutation effects predicted from sequence co-variation},
  volume   = {35},
  issn     = {1546-1696},
  doi      = {10.1038/nbt.3769},
  language = {eng},
  number   = {2},
  journal  = {Nature Biotechnology},
  author   = {Hopf, Thomas A. and Ingraham, John B. and Poelwijk, Frank J. and Schärfe, Charlotta P. I. and Springer, Michael and Sander, Chris and Marks, Debora S.},
  month    = feb,
  year     = {2017},
  pages    = {128--135}
}

@article{horta_ancestral_2022,
  title    = {Ancestral {Sequence} {Reconstruction} for {Co}-evolutionary models},
  volume   = {2022},
  issn     = {1742-5468},
  url      = {http://arxiv.org/abs/2108.03801},
  doi      = {10.1088/1742-5468/ac3d93},
  number   = {1},
  journal  = {Journal of Statistical Mechanics: Theory and Experiment},
  author   = {Horta, Edwin Rodríguez and Lage-Castellanos, Alejandro and Mulet, Roberto},
  month    = jan,
  year     = {2022},
  note     = {arXiv:2108.03801 [cond-mat]},
  pages    = {013502}
}

@article{jiang_learning_2022,
  title    = {Learning {Hyperbolic} {Embedding} for {Phylogenetic} {Tree} {Placement} and {Updates}},
  volume   = {11},
  issn     = {2079-7737},
  url      = {https://www.ncbi.nlm.nih.gov/pmc/articles/PMC9495508/},
  doi      = {10.3390/biology11091256},
  number   = {9},
  journal  = {Biology},
  author   = {Jiang, Yueyu and Tabaghi, Puoya and Mirarab, Siavash},
  month    = aug,
  year     = {2022},
  pages    = {1256}
}

@article{kalyaanamoorthy_modelfinder_2017,
  title      = {{ModelFinder}: fast model selection for accurate phylogenetic estimates},
  volume     = {14},
  copyright  = {2017 Springer Nature America, Inc.},
  issn       = {1548-7105},
  shorttitle = {{ModelFinder}},
  url        = {https://www.nature.com/articles/nmeth.4285},
  doi        = {10.1038/nmeth.4285},
  language   = {en},
  number     = {6},
  journal    = {Nature Methods},
  publisher  = {Nature Publishing Group},
  author     = {Kalyaanamoorthy, Subha and Minh, Bui Quang and Wong, Thomas K. F. and von Haeseler, Arndt and Jermiin, Lars S.},
  month      = jun,
  year       = {2017},
  pages      = {587--589}
}

@inproceedings{kingma_auto-encoding_2022,
  title     = {Auto-Encoding Variational Bayes},
  author    = {Kingma, Diederik P. and Welling, Max},
  booktitle = {Proceedings of the 2nd International Conference on Learning Representations (ICLR)},
  year      = {2014}
}

@article{levy_potts_2017,
  title    = {Potts {Hamiltonian} models of protein co-variation, free energy landscapes, and evolutionary fitness},
  volume   = {43},
  issn     = {0959-440X},
  url      = {https://www.ncbi.nlm.nih.gov/pmc/articles/PMC5869684/},
  doi      = {10.1016/j.sbi.2016.11.004},
  journal  = {Current opinion in structural biology},
  author   = {Levy, Ronald M and Haldane, Allan and Flynn, William F},
  month    = apr,
  year     = {2017},
  pages    = {55--62}
}

@book{liberles_ancestral_2007,
  title={Ancestral sequence reconstruction},
  author={Liberles, David A},
  year={2007},
  publisher={OUP Oxford},
  address = "New York"
}

@article{macaulay_fidelity_2023,
  title    = {Fidelity of hyperbolic space for {Bayesian} phylogenetic inference},
  volume   = {19},
  issn     = {1553-734X},
  url      = {https://pmc.ncbi.nlm.nih.gov/articles/PMC10166537/},
  doi      = {10.1371/journal.pcbi.1011084},
  number   = {4},
  journal  = {PLOS Computational Biology},
  author   = {Macaulay, Matthew and Darling, Aaron and Fourment, Mathieu},
  month    = apr,
  year     = {2023},
  pages    = {e1011084}
}

@inproceedings{mathieu_continuous_2019,
  title     = {Continuous Hierarchical Representations with {Poincar\'e} Variational Auto-Encoders},
  author    = {Mathieu, Emile and {Le Lan}, Charline and Maddison, Chris J. and Tomioka, Ryota and Teh, Yee Whye},
  booktitle = {Advances in Neural Information Processing Systems 32 (NeurIPS 2019)},
  pages     = {12565--12576},
  year      = {2019}
}

@article{matsumoto_novel_2021,
  title    = {Novel metric for hyperbolic phylogenetic tree embeddings},
  volume   = {6},
  issn     = {2396-8923},
  url      = {https://doi.org/10.1093/biomethods/bpab006},
  doi      = {10.1093/biomethods/bpab006},
  number   = {1},
  journal  = {Biology Methods and Protocols},
  author   = {Matsumoto, Hirotaka and Mimori, Takahiro and Fukunaga, Tsukasa},
  month    = jan,
  year     = {2021},
  pages    = {bpab006}
}

@article{mcgee_generative_2021,
  title    = {The generative capacity of probabilistic protein sequence models},
  volume   = {12},
  issn     = {2041-1723},
  url      = {https://www.nature.com/articles/s41467-021-26529-9},
  doi      = {10.1038/s41467-021-26529-9},
  language = {en},
  number   = {1},
  journal  = {Nature Communications},
  author   = {McGee, Francisco and Hauri, Sandro and Novinger, Quentin and Vucetic, Slobodan and Levy, Ronald M. and Carnevale, Vincenzo and Haldane, Allan},
  month    = nov,
  year     = {2021},
  pages    = {6302}
}

@article{muniz-trejo_epistatic_2025,
  title    = {Epistatic drift in protein evolution},
  volume   = {95},
  issn     = {0959-437X},
  url      = {https://www.sciencedirect.com/science/article/pii/S0959437X25001042},
  doi      = {10.1016/j.gde.2025.102412},
  journal  = {Current Opinion in Genetics \& Development},
  author   = {Muñiz-Trejo, Ricardo and Patton, Jaeda EJ and Herrera-Álvarez, Santiago and Thornton, Joseph W},
  month    = dec,
  year     = {2025},
  pages    = {102412}
}

@book{nei_molecular_2000,
  title={Molecular evolution and phylogenetics},
  isbn      = {978-0-19-988122-2},
  publisher = {Oxford University Press},
  author={Nei, Masatoshi and Kumar, Sudhir},
  address		= "New York",
  year      = "2000"
}

@inproceedings{nickel_poincare_2017,
  title     = {{Poincar\'e} Embeddings for Learning Hierarchical Representations},
  author    = {Nickel, Maximilian and Kiela, Douwe},
  booktitle = {Advances in Neural Information Processing Systems 30 (NeurIPS 2017)},
  pages     = {6338--6347},
  year      = {2017}
}

@article{nishikawa_epistasis_2021,
  title     = {Epistasis shapes the fitness landscape of an allosteric specificity switch},
  volume    = {12},
  copyright = {2021 The Author(s)},
  issn      = {2041-1723},
  url       = {https://www.nature.com/articles/s41467-021-25826-7},
  doi       = {10.1038/s41467-021-25826-7},
  language  = {en},
  number    = {1},
  journal   = {Nature Communications},
  publisher = {Nature Publishing Group},
  author    = {Nishikawa, Kyle K. and Hoppe, Nicholas and Smith, Robert and Bingman, Craig and Raman, Srivatsan},
  month     = sep,
  year      = {2021},
  pages     = {5562}
}

@article{paysan-lafosse_pfam_2025,
  title      = {The {Pfam} protein families database: embracing {AI}/{ML}},
  volume     = {53},
  issn       = {1362-4962},
  shorttitle = {The {Pfam} protein families database},
  url        = {https://doi.org/10.1093/nar/gkae997},
  doi        = {10.1093/nar/gkae997},
  number     = {D1},
  journal    = {Nucleic Acids Research},
  author     = {Paysan-Lafosse, Typhaine and Andreeva, Antonina and Blum, Matthias and Chuguransky, Sara Rocio and Grego, Tiago and Pinto, Beatriz Lazaro and Salazar, Gustavo A and Bileschi, Maxwell L and Llinares-López, Felipe and Meng-Papaxanthos, Laetitia and Colwell, Lucy J and Grishin, Nick V and Schaeffer, R Dustin and Clementel, Damiano and Tosatto, Silvio C E and Sonnhammer, Erik and Wood, Valerie and Bateman, Alex},
  month      = jan,
  year       = {2025},
  pages      = {D523--D534}
}

@article{rives_biological_2021,
  title     = {Biological structure and function emerge from scaling unsupervised learning to 250 million protein sequences},
  volume    = {118},
  url       = {https://www.pnas.org/doi/full/10.1073/pnas.2016239118},
  doi       = {10.1073/pnas.2016239118},
  number    = {15},
  journal   = {Proceedings of the National Academy of Sciences},
  publisher = {Proceedings of the National Academy of Sciences},
  author    = {Rives, Alexander and Meier, Joshua and Sercu, Tom and Goyal, Siddharth and Lin, Zeming and Liu, Jason and Guo, Demi and Ott, Myle and Zitnick, C. Lawrence and Ma, Jerry and Fergus, Rob},
  month     = apr,
  year      = {2021},
  pages     = {e2016239118}
}

@article{saitou_neighbor-joining_1987,
  title      = {The neighbor-joining method: a new method for reconstructing phylogenetic trees},
  volume     = {4},
  issn       = {0737-4038},
  shorttitle = {The neighbor-joining method},
  doi        = {10.1093/oxfordjournals.molbev.a040454},
  language   = {eng},
  number     = {4},
  journal    = {Molecular Biology and Evolution},
  author     = {Saitou, N. and Nei, M.},
  month      = jul,
  year       = {1987},
  pages      = {406--425}
}

@article{sennett_extant_2024,
  title      = {Extant {Sequence} {Reconstruction}: {The} {Accuracy} of {Ancestral} {Sequence} {Reconstructions} {Evaluated} by {Extant} {Sequence} {Cross}-{Validation}},
  volume     = {92},
  issn       = {0022-2844},
  shorttitle = {Extant {Sequence} {Reconstruction}},
  url        = {https://www.ncbi.nlm.nih.gov/pmc/articles/PMC10978691/},
  doi        = {10.1007/s00239-024-10162-3},
  number     = {2},
  journal    = {Journal of Molecular Evolution},
  author     = {Sennett, Michael A. and Theobald, Douglas L.},
  year       = {2024},
  pages      = {181--206}
}

@article{sgarbossa_generative_2023,
  title     = {Generative power of a protein language model trained on multiple sequence alignments},
  volume    = {12},
  issn      = {2050-084X},
  url       = {https://doi.org/10.7554/eLife.79854},
  doi       = {10.7554/eLife.79854},
  journal   = {eLife},
  publisher = {eLife Sciences Publications, Ltd},
  author    = {Sgarbossa, Damiano and Lupo, Umberto and Bitbol, Anne-Florence},
  editor    = {Colwell, Lucy J and Walczak, Aleksandra M},
  month     = feb,
  year      = {2023},
  pages     = {e79854}
}

@article{thornton_resurrecting_2004,
  title      = {Resurrecting ancient genes: experimental analysis of extinct molecules},
  volume     = {5},
  copyright  = {2004 Springer Nature Limited},
  issn       = {1471-0064},
  shorttitle = {Resurrecting ancient genes},
  url        = {https://www.nature.com/articles/nrg1324},
  doi        = {10.1038/nrg1324},
  language   = {en},
  number     = {5},
  journal    = {Nature Reviews Genetics},
  publisher  = {Nature Publishing Group},
  author     = {Thornton, Joseph W.},
  month      = may,
  year       = {2004},
  pages      = {366--375}
}

@article{wong_iq-tree_2025,
  title   = {{IQ-TREE} 3: phylogenomic inference software using complex evolutionary models},
  author  = {Wong, Thomas K. F. and Ly-Trong, Nhan and Ren, Huaiyan and Demotte, Piyumal and Ba{\~n}os, Hector and Roger, Andrew J. and Susko, Edward and Bielow, Chris and {De Maio}, Nicola and Goldman, Nick and Hahn, Matthew W. and {dos Reis}, Mario and Vinh, Le Sy and Huttley, Gavin and Lanfear, Robert and Minh, Bui Quang},
  journal = {Molecular Biology and Evolution},
  volume  = {43},
  number  = {5},
  pages   = {msag117},
  month   = may,
  year    = {2026},
  doi     = {10.1093/molbev/msag117},
  url     = {https://doi.org/10.1093/molbev/msag117}
}

@article{yule_iimathematical_1997,
  title     = {{II}.—{A} mathematical theory of evolution, based on the conclusions of {Dr}. {J}. {C}. {Willis}, {F}. {R}. {S}},
  volume    = {213},
  url       = {https://royalsocietypublishing.org/doi/10.1098/rstb.1925.0002},
  doi       = {10.1098/rstb.1925.0002},
  number    = {402-410},
  journal   = {Philosophical Transactions of the Royal Society of London. Series B, Containing Papers of a Biological Character},
  publisher = {Royal Society},
  author    = {Yule, George Udny},
  month     = jan,
  year      = {1925},
  pages     = {21--87}
}

@article{starr_pervasive_2018,
  author  = {Starr, Tyler N. and Flynn, Julia M. and Mishra, Parul and Bolon, Daniel N. A. and Thornton, Joseph W.},
  title   = {Pervasive contingency and entrenchment in a billion years of {Hsp90} evolution},
  journal = {Proceedings of the National Academy of Sciences},
  year    = {2018},
  volume  = {115},
  number  = {17},
  pages   = {4453--4458},
  doi     = {10.1073/pnas.1718133115}
}

@article{shah_contingency_2015,
  author  = {Shah, Premal and McCandlish, David M. and Plotkin, Joshua B.},
  title   = {Contingency and entrenchment in protein evolution under purifying selection},
  journal = {Proceedings of the National Academy of Sciences},
  year    = {2015},
  volume  = {112},
  number  = {25},
  pages   = {E3226--E3235},
  doi     = {10.1073/pnas.1412933112}
}

@article{poelwijk_learning_2019,
  author  = {Poelwijk, Frank J. and Socolich, Michael and Ranganathan, Rama},
  title   = {Learning the pattern of epistasis linking genotype and phenotype in a protein},
  journal = {Nature Communications},
  year    = {2019},
  volume  = {10},
  number  = {1},
  pages   = {4213},
  doi     = {10.1038/s41467-019-12130-8}
}

@article{morcos_direct-coupling_2011,
  author  = {Morcos, Faruck and Pagnani, Andrea and Lunt, Bryan and Bertolino, Arianna and Marks, Debora S. and Sander, Chris and Zecchina, Riccardo and Onuchic, Jos{\'e} N. and Hwa, Terence and Weigt, Martin},
  title   = {Direct-coupling analysis of residue coevolution captures native contacts across many protein families},
  journal = {Proceedings of the National Academy of Sciences},
  year    = {2011},
  volume  = {108},
  number  = {49},
  pages   = {E1293--E1301},
  doi     = {10.1073/pnas.1111471108}
}

@incollection{dayhoff_model_1978,
  author    = {Dayhoff, Margaret O. and Schwartz, Robert M. and Orcutt, Bruce C.},
  title     = {A model of evolutionary change in proteins},
  booktitle = {Atlas of Protein Sequence and Structure},
  editor    = {Dayhoff, Margaret O.},
  publisher = {National Biomedical Research Foundation},
  address   = {Washington, DC},
  year      = {1978},
  volume    = {5},
  pages     = {345--352},
  note      = {Supplement 3}
}

@article{jones_rapid_1992,
  author  = {Jones, David T. and Taylor, William R. and Thornton, Janet M.},
  title   = {The rapid generation of mutation data matrices from protein sequences},
  journal = {Computer Applications in the Biosciences},
  year    = {1992},
  volume  = {8},
  number  = {3},
  pages   = {275--282},
  doi     = {10.1093/bioinformatics/8.3.275}
}

@article{whelan_general_2001,
  author  = {Whelan, Simon and Goldman, Nick},
  title   = {A general empirical model of protein evolution derived from multiple protein families using a maximum-likelihood approach},
  journal = {Molecular Biology and Evolution},
  year    = {2001},
  volume  = {18},
  number  = {5},
  pages   = {691--699},
  doi     = {10.1093/oxfordjournals.molbev.a003851}
}

@article{le_improved_2008,
  author  = {Le, Si Quang and Gascuel, Olivier},
  title   = {An improved general amino acid replacement matrix},
  journal = {Molecular Biology and Evolution},
  year    = {2008},
  volume  = {25},
  number  = {7},
  pages   = {1307--1320},
  doi     = {10.1093/molbev/msn067}
}

@article{magee_robustness_2021,
  author  = {Magee, Andrew F. and Hilton, Sarah K. and DeWitt, William S.},
  title   = {Robustness of phylogenetic inference to model misspecification caused by pairwise epistasis},
  journal = {Molecular Biology and Evolution},
  volume  = {38},
  number  = {10},
  pages   = {4603--4615},
  year    = {2021},
  doi     = {10.1093/molbev/msab163}
}

@article{schmelkin_epistasis_2026,
  author  = {Schmelkin, Lisa and Chung, Sarah and Haldane, Allan and Townsend, Jeffrey P. and Carnevale, Vincenzo and Levy, Ronald M. and Kumar, Sudhir},
  title   = {Epistasis plays a limited role in driving entrenchment during neutral protein evolution},
  journal = {Genome Biology},
  volume  = {27},
  pages   = {272},
  year    = {2026},
  doi     = {10.1186/s13059-026-04157-0}
}

@article{felsenstein_evolutionary_1981,
  author  = {Felsenstein, Joseph},
  title   = {Evolutionary trees from {DNA} sequences: a maximum likelihood approach},
  journal = {Journal of Molecular Evolution},
  volume  = {17},
  number  = {6},
  pages   = {368--376},
  year    = {1981},
  doi     = {10.1007/BF01734359}
}

@article{yang_new_1995,
  author  = {Yang, Ziheng and Kumar, Sudhir and Nei, Masatoshi},
  title   = {A new method of inference of ancestral nucleotide and amino acid sequences},
  journal = {Genetics},
  volume  = {141},
  number  = {4},
  pages   = {1641--1650},
  year    = {1995},
  doi     = {10.1093/genetics/141.4.1641}
}

@inproceedings{sala_representation_2018,
  author    = {Sala, Frederic and De Sa, Chris and Gu, Albert and R{\'e}, Christopher},
  title     = {Representation tradeoffs for hyperbolic embeddings},
  booktitle = {Proceedings of the 35th International Conference on Machine Learning},
  series    = {Proceedings of Machine Learning Research},
  volume    = {80},
  pages     = {4460--4469},
  publisher = {PMLR},
  year      = {2018}
}

@inproceedings{sarkar_low_2011,
  author    = {Sarkar, Rik},
  title     = {Low distortion {Delaunay} embedding of trees in hyperbolic plane},
  booktitle = {Graph Drawing: 19th International Symposium, GD 2011},
  series    = {Lecture Notes in Computer Science},
  volume    = {7034},
  pages     = {355--366},
  publisher = {Springer},
  address   = {Berlin, Heidelberg},
  year      = {2011},
  doi       = {10.1007/978-3-642-25878-7_34}
}

@article{metropolis_equation_1953,
  author  = {Metropolis, Nicholas and Rosenbluth, Arianna W. and Rosenbluth, Marshall N. and Teller, Augusta H. and Teller, Edward},
  title   = {Equation of state calculations by fast computing machines},
  journal = {The Journal of Chemical Physics},
  volume  = {21},
  number  = {6},
  pages   = {1087--1092},
  year    = {1953},
  doi     = {10.1063/1.1699114}
}

@article{hastings_monte_1970,
  author  = {Hastings, W. K.},
  title   = {Monte {Carlo} sampling methods using {Markov} chains and their applications},
  journal = {Biometrika},
  volume  = {57},
  number  = {1},
  pages   = {97--109},
  year    = {1970},
  doi     = {10.1093/biomet/57.1.97}
}

@article{de_la_paz_epistatic_2020,
  author  = {de la Paz, Jose Alberto and Nartey, Charisse M. and Yuvaraj, Monisha and Morcos, Faruck},
  title   = {Epistatic contributions promote the unification of incompatible models of neutral molecular evolution},
  journal = {Proceedings of the National Academy of Sciences},
  volume  = {117},
  number  = {11},
  pages   = {5873--5882},
  year    = {2020},
  doi     = {10.1073/pnas.1913071117}
}

@article{bisardi_modeling_2022,
  author  = {Bisardi, Matteo and Rodriguez-Rivas, Juan and Zamponi, Francesco and Weigt, Martin},
  title   = {Modeling sequence-space exploration and emergence of epistatic signals in protein evolution},
  journal = {Molecular Biology and Evolution},
  volume  = {39},
  number  = {1},
  pages   = {msab321},
  year    = {2022},
  doi     = {10.1093/molbev/msab321}
}

@article{nee_tempo_1992,
  author  = {Nee, Sean and Mooers, Arne {\O}. and Harvey, Paul H.},
  title   = {Tempo and mode of evolution revealed from molecular phylogenies},
  journal = {Proceedings of the National Academy of Sciences},
  volume  = {89},
  number  = {17},
  pages   = {8322--8326},
  year    = {1992},
  doi     = {10.1073/pnas.89.17.8322}
}

@article{robinson_comparison_1981,
  author  = {Robinson, D. F. and Foulds, L. R.},
  title   = {Comparison of phylogenetic trees},
  journal = {Mathematical Biosciences},
  volume  = {53},
  number  = {1--2},
  pages   = {131--147},
  year    = {1981},
  doi     = {10.1016/0025-5564(81)90043-2}
}

@article{estabrook_comparison_1985,
  author  = {Estabrook, George F. and McMorris, F. R. and Meacham, Christopher A.},
  title   = {Comparison of undirected phylogenetic trees based on subtrees of four evolutionary units},
  journal = {Systematic Zoology},
  volume  = {34},
  number  = {2},
  pages   = {193--200},
  year    = {1985},
  doi     = {10.2307/2413326}
}

@article{bogdanowicz_matching_2012,
  author  = {Bogdanowicz, Damian and Giaro, Krzysztof},
  title   = {Matching split distance for unrooted binary phylogenetic trees},
  journal = {IEEE/ACM Transactions on Computational Biology and Bioinformatics},
  volume  = {9},
  number  = {1},
  pages   = {150--160},
  year    = {2012},
  doi     = {10.1109/TCBB.2011.48}
}

@article{smith_information_2020,
  author  = {Smith, Martin R.},
  title   = {Information theoretic generalized {Robinson}--{Foulds} metrics for comparing phylogenetic trees},
  journal = {Bioinformatics},
  volume  = {36},
  number  = {20},
  pages   = {5007--5013},
  year    = {2020},
  doi     = {10.1093/bioinformatics/btaa614}
}

@article{steel_distributions_1993,
  author  = {Steel, Michael A. and Penny, David},
  title   = {Distributions of tree comparison metrics---some new results},
  journal = {Systematic Biology},
  volume  = {42},
  number  = {2},
  pages   = {126--141},
  year    = {1993},
  doi     = {10.1093/sysbio/42.2.126}
}

@article{vinh_information_2010,
  author  = {Vinh, Nguyen Xuan and Epps, Julien and Bailey, James},
  title   = {Information theoretic measures for clusterings comparison: variants, properties, normalization and correction for chance},
  journal = {Journal of Machine Learning Research},
  volume  = {11},
  number  = {95},
  pages   = {2837--2854},
  year    = {2010}
}

@article{farris_estimating_1972,
  author  = {Farris, James S.},
  title   = {Estimating phylogenetic trees from distance matrices},
  journal = {The American Naturalist},
  volume  = {106},
  number  = {951},
  pages   = {645--668},
  year    = {1972}
}

@article{lupo_protein_2022,
  author  = {Lupo, Umberto and Sgarbossa, Damiano and Bitbol, Anne-Florence},
  title   = {Protein language models trained on multiple sequence alignments learn phylogenetic relationships},
  journal = {Nature Communications},
  volume  = {13},
  pages   = {6298},
  year    = {2022},
  doi     = {10.1038/s41467-022-34032-y}
}

@article{eddy_accelerated_2011,
  author  = {Eddy, Sean R.},
  title   = {Accelerated profile {HMM} searches},
  journal = {PLoS Computational Biology},
  volume  = {7},
  number  = {10},
  pages   = {e1002195},
  year    = {2011},
  doi     = {10.1371/journal.pcbi.1002195}
}

@inproceedings{clevert_fast_2016,
  author        = {Clevert, Djork-Arn{\'e} and Unterthiner, Thomas and Hochreiter, Sepp},
  title         = {Fast and accurate deep network learning by exponential linear units ({ELUs})},
  booktitle     = {International Conference on Learning Representations (ICLR)},
  year          = {2016},
  eprint        = {1511.07289},
  archivePrefix = {arXiv}
}

@article{srivastava_dropout_2014,
  author  = {Srivastava, Nitish and Hinton, Geoffrey and Krizhevsky, Alex and Sutskever, Ilya and Salakhutdinov, Ruslan},
  title   = {Dropout: a simple way to prevent neural networks from overfitting},
  journal = {Journal of Machine Learning Research},
  volume  = {15},
  number  = {56},
  pages   = {1929--1958},
  year    = {2014}
}

@inproceedings{ioffe_batch_2015,
  author    = {Ioffe, Sergey and Szegedy, Christian},
  title     = {Batch normalization: accelerating deep network training by reducing internal covariate shift},
  booktitle = {Proceedings of the 32nd International Conference on Machine Learning},
  series    = {Proceedings of Machine Learning Research},
  volume    = {37},
  pages     = {448--456},
  publisher = {PMLR},
  year      = {2015}
}

@inproceedings{miyato_spectral_2018,
  author        = {Miyato, Takeru and Kataoka, Toshiki and Koyama, Masanori and Yoshida, Yuichi},
  title         = {Spectral normalization for generative adversarial networks},
  booktitle     = {International Conference on Learning Representations (ICLR)},
  year          = {2018},
  eprint        = {1802.05957},
  archivePrefix = {arXiv}
}

@inproceedings{paszke_pytorch_2019,
  author    = {Paszke, Adam and Gross, Sam and Massa, Francisco and Lerer, Adam and Bradbury, James and Chanan, Gregory and Killeen, Trevor and Lin, Zeming and Gimelshein, Natalia and Antiga, Luca and Desmaison, Alban and K{\"o}pf, Andreas and Yang, Edward and DeVito, Zachary and Raison, Martin and Tejani, Alykhan and Chilamkurthy, Sasank and Steiner, Benoit and Fang, Lu and Bai, Junjie and Chintala, Soumith},
  title     = {{PyTorch}: an imperative style, high-performance deep learning library},
  booktitle = {Advances in Neural Information Processing Systems 32 (NeurIPS 2019)},
  pages     = {8024--8035},
  publisher = {Curran Associates, Inc.},
  year      = {2019}
}

@inproceedings{chadebec_pythae_2022,
  author        = {Chadebec, Cl{\'e}ment and Vincent, Louis J. and Allassonni{\`e}re, St{\'e}phanie},
  title         = {{Pythae}: unifying generative autoencoders in {Python} -- a benchmarking use case},
  booktitle     = {Advances in Neural Information Processing Systems 35 (NeurIPS 2022), Datasets and Benchmarks Track},
  publisher     = {Curran Associates, Inc.},
  year          = {2022},
  eprint        = {2206.08309},
  archivePrefix = {arXiv}
}

@inproceedings{nagano_wrapped_2019,
  author    = {Nagano, Yoshihiro and Yamaguchi, Shoichiro and Fujita, Yasuhiro and Koyama, Masanori},
  title     = {A wrapped normal distribution on hyperbolic space for gradient-based learning},
  booktitle = {Proceedings of the 36th International Conference on Machine Learning},
  series    = {Proceedings of Machine Learning Research},
  volume    = {97},
  pages     = {4693--4702},
  publisher = {PMLR},
  year      = {2019}
}

@article{blum_interpro_2025,
  author  = {Blum, Matthias and Andreeva, Antonina and Florentino, Laise Cavalcanti and Chuguransky, Sara Rocio and Grego, Tiago and Hobbs, Emma and Pinto, Beatriz Lazaro and Orr, Ailsa and Paysan-Lafosse, Typhaine and Ponamareva, Irina and Salazar, Gustavo A. and Bordin, Nicola and Bork, Peer and Bridge, Alan and Colwell, Lucy and Gough, Julian and Haft, Daniel H. and Letunic, Ivica and Llinares-L{\'o}pez, Felipe and Marchler-Bauer, Aron and Meng-Papaxanthos, Laetitia and Mi, Huaiyu and Natale, Darren A. and Orengo, Christine A. and Pandurangan, Arun P. and Piovesan, Damiano and Rivoire, Catherine and Sigrist, Christian J. A. and Thanki, Narmada and Thibaud-Nissen, Fran{\c{c}}oise and Thomas, Paul D. and Tosatto, Silvio C. E. and Wu, Cathy H. and Bateman, Alex},
  title   = {{InterPro}: the protein sequence classification resource in 2025},
  journal = {Nucleic Acids Research},
  volume  = {53},
  number  = {D1},
  pages   = {D444--D456},
  year    = {2025},
  doi     = {10.1093/nar/gkae1082}
}

\section{Acknowledgements}
This work is partially supported by the National Institutes of Health (R35GM139540-06). This research includes calculations carried out on HPC resources supported in part by the National Science Foundation through major research instrumentation grant number 1625061 and by the US Army Research Laboratory under contract number W911NF-16-2-0189. We acknowledge usage of Generative AI for certain parts of the code and text improvement. We would like to thank Ethan Klein, John Allard, and Rohan Alibutud for providing insights about the data gathering phylogenetic benchmarking.

\section{Contributions}

V.C. conceived the idea behind the project. J.L. and M.E.M. developed the model architecture. M.E.M. implemented core model functionality. J.L. contributed to model implementation, prepared the datasets, performed all analysis and wrote the manuscript. S.K., V.C., and X.S. supervised the project and revised the manuscript. 

\section{Competing interests}
The authors declare that they have no competing interests.

\end{document}